\documentclass[printer]{aa}
\usepackage{graphics,epsf,epsfig,graphicx}
\usepackage{natbib}
\usepackage{mathrsfs}
\usepackage{latexsym}
\usepackage{amssymb}
\usepackage{amsmath}
\usepackage{subfigure,rotating} 
\usepackage{url}
\usepackage{appendix}
\usepackage{listings}
\newcommand{\be}{\begin{equation}}
\newcommand{\ee}{\end{equation}}

\renewcommand{\arraystretch}{2.}
\begin{document}

\title{Reionization by UV or X-ray sources}

\author{S. Baek\inst{1,2}, B. Semelin\inst{1,3}, P. Di Matteo\inst{4}, Y. Revaz\inst{5}, F. Combes\inst{1}}
\offprints{sunghye.baek@sns.it}
\institute{
LERMA, Observatoire de Paris, 61 Av. de l'Observatoire, F-75014, Paris, France
\and
Scuola Normale Superiore, Piazza dei Cavalieri 7, 56126 Pisa, Italy
\and
Universit\'e Pierre et Marie Curie, 4 place Jussieu 75005 Paris, France
\and
GEPI, Observatoire de Paris, 5, place Jules Janssen 92195 Meudon Cedex, France
\and
Laboratoire d'Astrophysique, \'Ecole Polytechnique F\'ed\'erale de Lausanne (EPFL), Swizerland
}
\date{Received 2010 Mars 3; accepted 2010 August 3 }

\abstract{
We present simulations of the 21-cm signal 
during the epoch of reionization. We focus on properly modeling the absorption
regime in the presence of inhomogeneous Wouthuysen-Field effect and X-ray heating.
We ran radiative transfer simulations for three bands in the source spectrum
(Lyman, UV, and X-ray) to fully account for these processes. 
We find that the brightness temperature fluctuation of the 21 cm signal has an
 amplitude greater than 100 mK during the early reionization, up to 10 times greater
than the typical amplitude of a few 10 mK obtained during the later emission phase. 
More importantly, we find that even a rather high contribution from QSO-like sources
only damps the absorption regime without erasing it. Heating the IGM with X-ray
takes time.  Our results show that observations of the early reionization will probably
benefit from a higher signal-to-noise value than during later stages. After
analyzing the statistical properties of the signal (power spectrum and PDF) we find
three diagnostics to constrain the level of X-ray, hence the nature of the
first sources.
}

\maketitle

\section{Introduction}

The epoch of reionization (EoR) started with the formation of the first sources of light around $z=15 - 30$. As shown
by the Gunn-Peterson effect \citep{Gunn65} in the spectra of high-Fredshift quasars (QSO) (e.g., 
\citealt{Fan06}), the universe was fully reionized by $z \sim 6$. WMAP 5-year results show that the optical depth
 for the Thomson scattering of 
CMB photons traveling through the reionizing universe is $\tau= 0.084 \pm 0.016$ \citep{Koma08}. Together
with the Gunn-Peterson data, this strongly favors an extended reionization period between $z>11$ and $z=6$. 

While other observations,
such as the Lyman-$\alpha$ emitter luminosity function \citep{Ouch09}, may produce other constraints on the history of
reionization in the next few years, the most promising is the observation of the $21$-cm line in the neutral 
IGM using large radio-interferometers 
(LOFAR\footnote{LOw Frequency ARray, \url{http://www.lofar.org}}, 
MWA\footnote{Murchison Widefield Array, \url{http://mwatelescope.org/}}, 
GMRT\footnote{Giant Metrewave Radio Telescope, \url{http://www.gmrt.ncra.tifr.res.in}},
21-CMA\footnote{21 Centimeter Array, \url{http://21cma.bao.ac.cn/}}, SKA\footnote{Square Kilometre Array, 
\url{http://www.skatelescope.org/}}). 
The signal will be observed in emission or in absorption against the CMB continuum. Both theoretical modeling 
\citep{Mada97,Furl06a} and simulations (e.g., \citealt{Ciar03c,Gned04,Mell06b,Lidz08,Ichi09,Thom09}) 
 show that the brightness temperature fluctuations of the 21 cm signal have an 
amplitude of a few $10$ mK in emission, on scales from tens of arcmin down to sub-arcmin. 
With this amplitude, and ignoring the issue of
foreground cleaning residuals, statistical quantities such as the three-dimensional 
power spectrum should be measurable with LOFAR or MWA with a few
100 hours integration \citep{Mora04,Furl06a,Lidz08}.
In absorption however, the amplitude of the fluctuations 
may exceed $100$ mK \citep{Gned04,Sant08,Baek09}, the exact
level depending on the relative contribution of the X-ray and UV sources to the process 
of cosmic reionization.

 The signal will be seen in
absorption during the initial phase of reionization, probably at $z> 10$, 
when the accumulated amount of emitted X-ray is not yet sufficient to raise the
IGM temperature above the CMB temperature. The duration and intensity of this 
absorption phase, regulated by the spectral energy distribution 
(SED) of the sources, are crucial. SKA precursors able to probe the relevant frequency 
range, 70 - 140 MHz, may benefit from a much higher
signal-to-noise than during later periods in the EoR. However, if the absorption phase is
 confined at redshifts above $15$, RFI and the ionosphere
will become an problem. Quite clearly, the different types of sources of reionization and different 
formation histories produce very different properties
for the 21-cm signal. It is therefore important for future observations to explore the range of astrophysically plausible
scenarios using numerical simulations. 

To properly model the signal, it is necessary to use $ > 100 {h}^{-1}\mathrm{Mpc}$ 
box sizes \citep{Bark04}. Together with
a large box size, it is desirable to resolve halos with masses down to $10^8 M_{\odot}$ 
as these contain sources (able to cool below their virial temperature by atomic processes), 
or even minihalos with masses down
to $10^4 M_{\odot}$ became they act as an efficient photon sink because of their high recombination 
rate \citep{Ilie05}. As this work focuses on improving the physical modeling,
we restrict ourselves to resolving halos with a mass $10^{10} M_{\odot}$ or higher. 
Indeed, simulating the absorption phase correctly, as we do in this work, requires a more 
extensive and more costly implementation of radiative transfer. We are exploring the direct
implication of this improved physical modeling, and will turn to better mass resolution
in the near future.

There are three bands in the sources SED that influence the level of the 21 cm signal: 
the Lyman band, the ionizing UV band, and the soft X-ray band.
Lyman band photons are necessary to decouple the spin temperature of hydrogen from the CMB 
temperature through the Wouthuysen-Field effect \citep{Wout52,Fiel58},
 and make the EoR signal visible. UV band photons 
are of course responsible for the ionization of the IGM, and
soft X-rays are able to preheat the neutral gas ahead of the ionizing front, deciding whether 
the decoupled spin temperature is less (weak preheating) or
greater (strong preheating) than the CMB temperature. While a proper modeling should perform the 
full 3D radiative transfer in all 3 bands, a simpler modeling
has often been used in previous works. Indeed, for the usual source SEDs and source formation
 histories, once the average ionization fraction of the universe is greater 
than $\sim 10 \%$, the background flux of Lyman-$\alpha$ photons is so high that the hydrogen 
spin temperature is fully coupled to the kinetic temperature
by the Wouthuysen-Field effect \citep{Baek09}. Thereafter, computing the Lyman band 
radiative transfer is unnecessary. In the same spirit,
it has usually been assumed that the preheating of the IGM by soft X-ray was strong enough to raise 
the kinetic temperature much higher than the CMB 
temperature everywhere early in the EoR. However, both assumptions fail during the early reionization:
the absorption phase. 
Even in the later part of reionization the second assumption may fail, depending on the nature of the
 sources. We will quantify this possibility in this paper.
Computing the full radiative transfer in all three bands is necessary to study the absorption regime.
 Indeed, fluctuations in the local Lyman-$\alpha$ flux
induce fluctuations in the spin temperature (while the Wouthuysen-Field effect is not yet saturated),
 which, in turn, modify the power spectrum of the
$21$ cm signal \citep{Bark05, Seme07, Chuz07, Naoz08, Baek09}.
The same is true for the fluctuations in the local flux of X-ray photons \citep{Prit07,Sant08}.

Let us emphasize however that, in modeling Lyman-$\alpha$ and X-ray fluctuations,
\citet{Bark05}, \citet{Naoz08}, \citet{Prit07} and
\citet{Sant08} all use the semi-analytical approximation that the IGM has a uniform density
 of absorbers and ignore wing effects in the radiative transfer of Lyman-$\alpha$
photons.  \citet{Seme07} and \citet{Chuz07} have shown that 
these wing effects do exist. Moreover, once reionization is under way,
ionized bubbles create sharp fluctuations in the number density of absorbers (not to mention simple 
matter density fluctuations). In this work, for the first time,
we present results based on simulations with full radiative transfer for both Lyman-$\alpha$ and X-ray photons.

What are the possible candidates as sources of reionization? Usually, two categories are considered:
 ionizing UV sources (Pop II and III stars), and X-ray sources (quasars).
When we study 21 cm absorption, however, we must distinguish between Pop II and Pop III stars beyond 
the large difference in luminosity per
unit mass of formed star. Indeed Pop II stars have a three times larger Lyman band to ionizing UV band
 luminosity ratio than Pop III stars.
 It means that the 21 cm signal
will reach its full power (near saturated Wouthuysen-Field effect) at a lower average ionization 
fraction for Pop II stars than for Pop III stars. The relevant
question is: how long do Pop III stars dominate the source population before Pop II stars take over? 
The answer to this question, related to the whole process
of star formation, feedback and metal enrichment of the IGM, is a difficult one. At this stage,
 state of the art numerical simulations of the EoR use simple
prescriptions in the best case (e.g. \citealt{Ilie07a}), or simply ignore this issue. 

 The other category of sources are X-ray sources. They may be 
mini-quasars, X-ray binaries, supernovae \citep{Oh01, Glov03}, or even
 more exotic candidates such as dark stars 
\citep{schl09}. The exact level of emission from these sources is 
a matter of speculation. The generally accepted view is that stars
dominate over X-ray sources and are sufficient to drive reionization 
\citep{Shap87, Giro96, Mada99, Ciar03a}. Recently, \citet{Volo09} supported
the opposite view. While, in their models, X-ray sources are marginally able to complete
reionization by $z \sim 6$, they find a very low contribution from stars. Indeed they rely
on \citet{Gned08} who find, using numerical simulations, a negligible escape
fraction for ionizing radiations from galaxies with total mass less than a few $10^{10} M_{\odot}$, who
should actually contribute to $90\%$ of the ionizing photon production during the EoR \citep{Chou07}.
While the physical modeling in their innovative simulations is quite detailed, this surprizing behavior 
of the escape fraction definitely needs to be checked at higher resolution and with different codes.
For the time being the best simulations can only explore a plausible range of X-ray contributions, and quantify
the impact on observables. When the observations become available we would like to be able, using 
simulation results, to derive tight constraints on the relative 
level of emission from ionizing UV and X-ray sources. This work, exploring the 21 cm signal for a
 few different levels of X-ray emission, is a first step toward this
goal.  

The paper is organized as follows. We present the numerical methods in \S2 and
describe our source models in \S3. 
In \S4, we show the results and analyze the differences between the
models. We discuss our findings and conclude in \S5.

\section{Numerical simulation}
 The numerical methods used in this work are similar to those
presented in \citet{Baek09} (hereafter Paper I). The references to previous and some new validation
tests are presented in the Appendix.
The dynamical simulations have been run with GADGET-2 \citep{Spri05}
and post-processed with UV continuum radiative transfer and further processed with Ly-$\alpha$
transfer using LICORICE.
The same cosmological parameters and particle number are used
and  we refer the reader to Paper I
for details related to the numerical methods and parameters.
The main improvement on the previous work is using a more realistic source model including
soft X-ray and implementing He chemistry. 

We have run seven different simulations, all of which use the same $100\,h^{-1}$Mpc box, density
fields, and star formation rate, but with different initial mass functions (IMF), chemistry (with helium or without),
X-ray fraction of the total luminosity or X-ray spectral index.
S1 is the reference model. S2 has a top-heavy IMF (Salpeter IMF restricted to a $100-120 M_\odot$ range), while the others uses a Salpeter IMF in a $1.6-120 M_\odot$ range.
Only S3 contains helium. In all other models, helium is replaced by the same mass of hydrogen. X-ray radiative transfer is included in S4, S5, S6 and S7.
They have either different X-ray fraction of the total luminosity or X-ray spectral index.
The basic parameters of these simulations are summarized in Tab.~\ref{model}

The simulations are controlled by a few parameters. 
We adopted the same value as in paper I for the maximum value of the number of particles per radiative transfer cell in the adaptive grid: $N_{max}=30$. The resulting minimum radiative transfer cell size is 
200 $h^{-1}$ kpc at $z=6.6$.
Between two snapshots, i. e. $\sim 10$ Myr, we cast $3 \times 10 ^{6}$ photon packets for photoionization (all in the UV for models S1 to S3, half in the UV and half X-rays for models S4 to S7), and
$3 \times 10^{7}$ photons for Lyman-$\alpha$ transfer. At the end of the simulations ($z \sim 6$), the number of sources
reaches $\sim 15000$, so the number of ionizing photon packets per source is only 200.
However, at this final stage the sources are highly clustered and very large and ionized regions
surround the source clusters. So the clustered sources cooperate to reduce the Monte Carlo noise at the
ionization fronts.
In addition, the adaptive grid responds better than a fixed grid to sampling issues: big cells where there
 are few photons, small cells where there are many. \citet{Mase05} presents convergence tests for a 
Monte-Carlo radiative transfer code very similar to ours. Their convergence tests suggest that the typical 
level of noise in our ionization and temperature cubes is $\sim 10\%$. We accept is as a reasonable value, 
especially since, having run the \citet{Ilie09} comparison tests, we are confident that our ionization 
fronts propagate at the correct speed.
We use 1000 frequency bins in each of the photoionizing-UV and X-ray spectra. For Lyman-$\alpha$ transfer,
we sample the frequency at random between Lyman-$\alpha$ and Lyman-$\beta$.

\begin{table}[hbp]
\centering
\small
\tabcolsep 3pt
\renewcommand\arraystretch{1.2}
   \begin{tabular}{c c c c c c}
\hline
\hline
    Model   &IMF & Helium & $L_{\rm{star}}$ &$L_{\rm{QSO}}$  & spectral index     \\
\hline
\hline
    S1      &$1.6-120M_{\odot}$  & No        & 100 \%        &0 \%    &-    \\
    S2      &$100-120M_{\odot}$  & No            & 100 \%        &0 \%    &-     \\
    S3     &$1.6-120M_{\odot}$   & Yes         &100 \%         &0 \%      &- \\
    S4     &$1.6-120M_{\odot}$   & No        &99.9 \%        &0.1 \%      &$\alpha$=1.6 \\
    S5      &$1.6-120M_{\odot}$  & No        &99.9 \%        &0.1 \%     &$\alpha$=0.6 \\
    S6     &$1.6-120M_{\odot}$   & No        &99 \%        &1 \%      &$\alpha$=1.6 \\
    S7     &$1.6-120M_{\odot}$   & No        &90 \%        &10 \%      &$\alpha$=1.6 \\
\hline
\end{tabular}
\caption[]{Simulation parameters. $L_{star}$ is the stellar luminosity fraction
and $L_{QSO}$ is the X-ray luminosity fraction of the total luminosity.}
\label{model}
\end{table}

 \subsection{X-ray radiative transfer}
\label{X-ray}
The main difference between the cosmological radiative transfer of ionizing UV and X-ray is the mean free
path of the photons, at most a few $10$ comoving Mpc in the first case, possibly several $100$ Mpc
in the second case. A usual
trick in implementing UV transfer is to use an infinite speed of light: do so with LICORICE (see Paper I). This is correct
if the crossing time of the photons between emission and absorption points is much less than the recombination time,
the photo-ionization time \citep{Abel99} and the typical time for the variation of luminosity of the sources. This is the 
case in most of the IGM during the EoR, except very close to the sources where the photo-ionizing rate is very high. 
Obviously, this is not the case for X-rays which have a much longer
crossing time. Consequently, we implemented the correct propagation speed for X-ray photons. 
We propagate an X-ray photon packet during one radiative transfer time step $\Delta t_{reg}$ ($< 1$ Myr, same notation
from Fig.1 of Paper I) 
over a distance of $c \Delta t_{reg}$, where $c$ is the velocity of light.
 Then, the frequency of the photon packet and photoionizing cross sections are 
recomputed with the updated value of the cosmological expansion factor.  
The photon packet propagates during the next radiative transfer time step using these new
parameters.
If the photon packet loses 99\% of its initial energy, we drop it.
X-ray photon packets containing photons with an energy of
 several keV pass through the periodic simulation box several times
before they lose most of their initial content. For each density snapshot, that is every $\sim 10$ Myr, 
1.5 millions of photon packets are sent from the X-ray sources. About half of them are  absorbed during the
computation on the same density snapshot when they were emitted,
 and the other half is stored in memory to be propagated through the next density snapshots.
 Some X-ray packets with very high energy photons still survive several snapshots later, so the 
number of stored photon packets grows as simulations progress.
About 50 millions photon packets are stored in memory toward the end of the simulations.

It may seem that this memory overhead, which sets a limitation to the possible simulations with LICORICE,
 would not appear with  radiative transfer algorithms which naturally include a finite velocity
of light like moment methods. However, these methods would suffer from an overhead connected to the
number of frequency bins necessary to correctly model X-rays, while it does not exist in Monte-Carlo
methods. Including complete X-ray transfer in EoR simulations comes at a non-negligible
cost, whatever the numerical implementation. 
 Since the X-ray photons can propagate over several box sizes during several tens of Myr,
the X-ray frequency can redshift considerably between emission and absorption.
The cross-section of photoionization has a strong frequency dependence, so we
have to redshift the frequency of the photons. At each radiative transfer
time step $\Delta t_{reg}$, we update the frequency of all the X-ray photon packets,

\be
\nu (t+\Delta t_{reg})=\frac{a(t)}{a(t+\Delta t_{reg})}\nu(t),
\ee
where $a(t)$ is the expansion factor of the Universe.

The treatment of non-thermal electrons produced by X-ray will be described in \S \ref{sourceX}

 \subsection{Helium reionization}
The intergalactic medium is mainly composed of hydrogen and helium, with contributions of
90\% and 10\% in number. Until now, we have run simulations with hydrogen only, but
including helium is worth studying because the different value of the ionization thresholds and
photoionization rates could affect the reionization history. 
We included He, $\text{He}^+$, and $\text{He}^{++}$ in LICORICE, and used
\citet{Cen92} and \citet{Vern96} for various cooling rate and cross sections.
When helium chemistry is turned on, the ionization fractions ($\text{H}^+$, $\text{He}^+$ and 
$\text{He}^{++}$) and the temperature are integrated explicitly using the adaptive scheme described 
in Paper I. More details on the numerical methods and a validation tests of the treatment of helium are presented in Appendix.

\section{Source model}

  \subsection{Computing the star formation rate}

Our new source model needs the star formation rate for all baryon particles.
We recompute the star formation in the radiative transfer
simulations rather than to rerun the dynamical simulation. Here is why and how.

We adopted the procedure described in \citet{Miho94}, employing a “local”
 Schmidt law and an hybrid-particles algorithm to implement it
in our code.
 Indeed, in our model,
 the star formation rate  solely depends on the density, and we make the assumption 
that the star formation feedback (kinetic and thermal) on the dynamics does not vary much 
from the fiducial simulation.
LICORICE uses the classical Schmidt law:
\be
\frac{df _*}{dt}=\frac{1}{t_*} \,\,\,(\text{if}\,\,\,\rho_{g} > \rho_{\text{threshold}}) ,
\ee
where $t_*$ is defined by :
\be
t_*=t_{0*}\left( \frac{\rho_g}{\rho _{\text{threshold}}} \right)^{-1/2} .
\ee
$\rho _g$ is the gas density and $f _*$ is the star fraction.  
We set the parameters $t_{0*}$ and $\rho _{\text{threshold}}$  so that the evolution of the
global star fraction follows closely that of the S20 simulation ($20 h^{-1}$ Mpc) in Paper I, and 
reionization completes at $z \sim 6$. In this way, we reuse the tuning made for the S20 simulation, and
at high $z$, we get a similar star formation history as in the higher resolution (but smaller box size) S20
simulation.
All simulations in the present work have a $100 h^{-1}$ Mpc box size.
Following the above equations, a gas particle whose local density exceeds the threshold
($\rho_{\text{threshold}} =5\,\rho _{\rm{critical}}\times \Omega_{b}$) increases its star fraction, $f_*$, where
$\rho _{\rm{critical}}$ is the critical density of the universe and $\Omega_{b}$ is the cosmological
 baryon density parameter.

\subsection{Limiting the number of sources}

 We compute the \emph{increase} in the star fraction for each particle, 
$\Delta f_*$ between two consecutive snapshots.
Then the total mass of young stars formed in a particle is 
$m \times \Delta f_*$, where $m$ is the mass of the particle.
To avoid a huge number of sources, we had to
set a threshold on the new star fraction for the particle to act as a source. 
We used $\Delta f_*>0.001$. We checked that this leaves out a negligible amount of star formation,
about 0.4\%.  It happens that several source particles reside in the same radiative transfer cell, but we treated individually
 ray tracing for each source.

\subsection{Choosing an IMF}
 With our mass resolution, this amount $m_{\text{gas}} \times \Delta f_*$ corresponds
to a star cluster or a dwarf galaxy so an IMF should be taken into account.
We choose a Salpeter IMF, with masses in the range $1.6\text{M}_{\odot}-120\text{M}_{\odot}$ or
$100\text{M}_{\odot}-120\text{M}_{\odot}$ (model S2). The first range is used to model 
the SED of an intermediated Pop II and Pop III star population, and the other one is for pure Pop III stars.

 \subsection{Computing the luminosity and SED of the stellar sources}

The next step is to make the link between the amount of created stars and the luminosity of the 
sources. When only the ionizing UV luminosity is considered, it is quite justified to use simple   
models. For example we can make it proportional to the mass of the host dark matter halo \citep{Ilie06b},
or, as we did in Paper I, to the mass of the baryonic particles newly converted into stars. Things
are more complicated when we consider both the Lyman band and ionizing UV. Indeed, since each particle
is massive enough to contain a representative sample of the choosen IMF, and since
each mass bin has a different life time, we should consider an SED evolving with the age of the star particle.
This would be possible using pre-tabulated SEDs. However, unlike in Paper I, we decided to use hybrid particles 
which begin to produce photons as soon as a small fraction of the particle is turned into stars. This is
useful to make the local luminosity less noisy in the early EoR when the source mass resolution is an issue.
Including this star formation history
for each particle and convolving with the time-varying SED would be extremely costly in terms of both memory
and computation time.

 We simplified the issue by considering the fact that in the Lyman and UV band, most of the luminosity
is produced by the massive stars, with a short life time comparable to the time between two snapshots of 
our simulations. So we decided to use a constant SED and luminosity 
during a characteristic life time. Both luminosity and SED are computed independently in each of the Lyman and 
ionizing band. To compute the luminosity and the SED for a star particle we use the data for massive, low 
metallicity ($Z=0.004Z_{\odot}$) stars in the main sequence \citep{Meyn05,Hans94} (see Tab.~\ref{star_info}).
The details of how this is done can be found in Appendix.
The constant luminosity and characteristic life time, computed in the two spectral bands, are given in Tab.~\ref{luminosity_life} 
We find characteristic life time of $< 8$ Myr for the UV band.  In the implementation of the UV transfer however, for technical reasons,
the source fraction of the particle actually shines for a duration equal the interval between two snapshots. This varies 
varies between $6$ and $20$ Myr, so we recalibrate the luminosity to produce the correct amount of energy. The whole
point of the procedure, is to take the different typical life time in the Lyman band into account, especially at at $z$,
when Lyman-$\alpha$ coupling is not yet saturated. We should not concentrate the emission within a single
snapshot interval, which is 3 times shorter than the source life-time, or we would artificially boost the coupling between t
he spin temperature of hydrogen and the kinetic temperature
of the gas and alter the resulting brightness temperature. Consequently we let each newly formed star fraction of a particle shine
for $3$ consecutive snapshots, which is close to the typical life time in the Lyman band, and we still recalibrate the luminosity 
to produce the correct amount of energy. While we do not
use a time-evolving SED, we believe that implementing different life times for the Lyman and UV sources with the correctly
average luminosities is a substantial improvement in our source model.

We use an escape fraction $f_{esc}=0.12$ for photoionizing UV photons and $f_{esc}=1$ for Lyman-$\alpha$.

\begin{table}
\centering
\tabcolsep 5.8pt
\renewcommand\arraystretch{1.2}
   \begin{tabular}{ c c c c}
\hline
\hline
     mass $[M_{\odot}]$    & log$(L/L_{M_{\odot}})$ & log$(T_{\mathrm{eff}})$&$t_{\mathrm{life}} [Myr]$    \\
\hline
\hline
    120        & 6.3    & 4.7    & 3  \\
    60        & 5.8    & 4.6    & 4.5 \\
    40        & 5.6    & 4.5    & 6 \\
    30        & 5.2    & 4.5    & 7 \\
    20        & 4.8    & 4.45    & 10 \\ 
    15        & 4.65    & 4.4    & 14 \\
    12        & 4.2    & 4.37    & 20 \\
    9        & 3.8    & 4.3    & 34 \\
    5.9        & 2.92    & 4.18    & 120 \\
    2.9        & 1.73    & 3.97    & 700 \\
    1.6        & 0.81    & 3.85    & 3000 \\
    
\hline
\end{tabular}
\caption[]{Physical properties of low metalicity ($Z=0.004Z_{\odot}$) main sequence stars \citep{Meyn05}
L is the bolometric luminosity, $(T_{\mathrm{eff}})$ is the effective temperature and $t_{\mathrm{life}} [Myr]$ is the life time
of the star. }.
\label{star_info}
\end{table}

\begin{figure}[h]
 \centering
 \resizebox{\hsize}{!}{\includegraphics{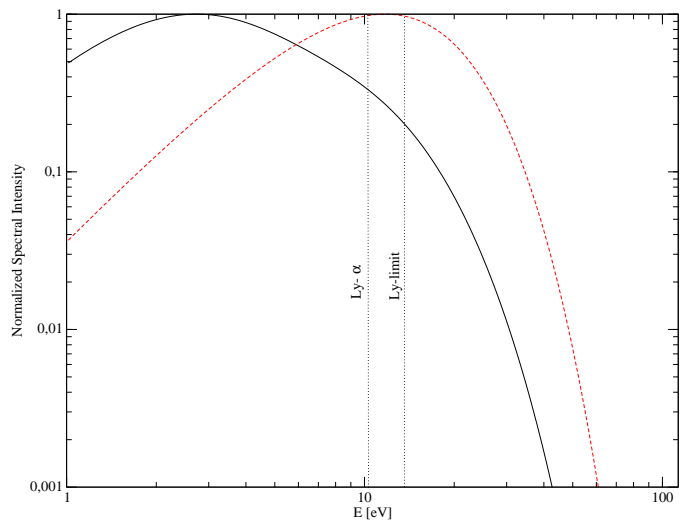}}
 \caption{Normalized spectral intensity of our source model. Black solid line is the SED from Salpeter IMF
and red dashed line is from top-heavy IMF.}
\label{SED}
 \end{figure}

\begin{table*}
\centering
\renewcommand\arraystretch{1.2}
   \begin{tabular}{ c c c c }
\hline
\hline
   IMF mass range  &    Energy band       & $10.24\, \rm{eV}  \leqslant E < 13.6 \,\rm{eV}$        & $E \geqslant 13.6 \,\rm{eV}$     \\
\hline
\hline
  1-120 $[M_{\odot}]$ &  Luminosity[erg/s]$^A$            & $6.32 \times 10^{44} $            & $2.14 \times 10^{45}$    \\
   &  Life time[Myr]$^A$        & $20.36$            & $8.03$     \\

\hline

  100-120 $[M_{\odot}]$ & Luminosity[erg/s]$^B$            & $9.96 \times 10^{45} $            & $3.12 \times 10^{46}$    \\
  &  Life time[Myr]$^B$        & $3.32$            & $3.31$     \\

\hline
\end{tabular}
\caption[]{Averaged luminosities and life times of our source model for a baryon particle 
depending on the energy band.
A values are from Salpeter IMF and B values are form top-heavy IMF. }
\label{luminosity_life}
\end{table*}

  \subsection{X-ray source model}
\label{sourceX}
X-rays can have a significant effect on the 21 cm brightness temperature.
The X-ray photons, having a smaller ionizing cross-section, can penetrate
neutral hydrogen further than UV photons and heat the gas above the CMB 
temperature. This X-ray heating effect on the IGM is often assumed to be homogeneous
because of X-rays' long mean free path. In reality, the X-ray flux is stronger around the
sources and the inhomogeneous X-ray flux can bring on extra fluctuations for the 21 cm brightness
temperature \citep{Prit07,Sant08}. Moreover, patchy reionization induce further fluctuations in the
local X-ray flux which can only be accounted for using a full radiative transfer modeling.

\subsubsection*{X-rays luminosity}
First, we need to determine the luminosity and location of X-rays sources.
We simply divided the total luminosity, $L_{tot}$, of all source particles
into a stellar contribution, $L_{star}$,
and a QSO contribution, $L_{QSO}$. 
$L_{QSO}$ depends on the star formation rate, since  $L_{tot}$ itself is proportional to
the increment of the star fraction, $\Delta f_*$, between two snapshots of the dynamic simulation.

One reason for this approach is that X-ray binaries and supernova remnants
contribute to X-ray sources as well as quasars
and they are strongly related to the star formation rate.
Following the work of \citet{Glov03}, we took 0.1\% of $L_{tot}$ as the fiducial X-ray source
luminosity, $L_{QSO}$. However, considering that they assumed a simple and empirically motivated model
we have also run simulations with different values $L_{QSO}$, 1\% and 10\% of $L_{tot}$.
 Quasar luminosity fractions less than 0.1\% are not of interest for us, since
their heating effect will be negligible.

\subsubsection*{X-ray energy range and nature of the sources}
First, we have to choose the photon energy range since hard X-ray photons have
a huge mean free path which costs a lot for ray-tracing computations.
The comoving mean free path of an X-ray with energy $E$ is \citep{Furl06b}
\be
\lambda _{X}=4.9 \overline{x} ^{-1/3} _{\rm{HI}} \left( \frac{1+z}{15}\right) ^{-2}
\left( \frac{E}{300eV} \right) ^3 \,\,\rm{comoving\ Mpc}.
\ee
Only photons with energy below $E\sim 2[(1+z)/15]^{1/2}\overline{x}^{1/3}_{\rm{HI}}\,k$eV
are absorbed within a Hubble time and the $E^{-3}$ dependence of the cross-section means that heating
is dominated by soft X-rays, which do fluctuate on small scales (Furlanetto et al. 2006). Therefore,
we choose an energy range for X-ray photons from $0.1k$eV to $2k$eV. The photons with energy higher
than $2k$eV are not absorbed until the end of simulation at $z\approx 6$.

 While the most likely astrophysical sources of X-ray during the EoR are supernovae, X-binaries and (mini-) quasars,
it is interesting to mention that the X-ray SED of supernovae and X-binaries typically peaks above $1$ $k$eV (e.g. \citealt{Oh01}).
This means that most of the X-rays emitted by these sources will interact with the IGM more than $10^8$ years later, which is not true
for QSO-like SEDs. During this time interval the global source mass (and, to first order, luminosity) easily rises by a factor of 10.
Thus the longer delay will lower the effective luminosity of X-binaries and supernovae compared to QSO. For this reason, but also to
avoid detailed modeling of some aspects while others, like the overall luminosity of X-ray sources, remain largely unconstrained, we
use QSO as our typical X-ray source.

\subsubsection*{QSO spectral Index}
We model the specific luminosity of our QSO-like sources as a
power-law  with index $\alpha$;
\be
L_{\nu}=k\left( \frac{\nu}{\nu_0} \right)^{-\alpha}.
\label{power_index}
\ee
 $k$ is a normalization constant so that
\be
L_{QSO}=\int ^{\nu _2} _{\nu _1} L_{\nu} \,\rm{d}\nu ,
\label{power_index2}
\ee
where $h\nu _1=0.1\,k$eV and $h\nu _1=2\,k$eV.
The amount of X-ray heating can be altered by the shape of the spectrum but
there exists a large observational uncertainty in the mean and distribution of $\alpha$.
We extrapolate from the measurement of extreme UV spectral index by \citet{Telf02} to higher energy.
The index values are derived from fitting $1 Ry < E < 4 Ry$, and we extrapolated to 2$k$eV.
The measured value by \citet{Telf02} is approximately
$\approx$ 1.6, but with a large gaussian standard deviation of 0.86.
\citet{Scot04} derived an average value of $\alpha=0.6$ from a sample of FUSE and HST quasars.
We choose $\alpha=1.6$ as our fiducial case, and used $\alpha=0.6$ for comparison.

\subsubsection*{Secondary Ionization}
 X-rays deposit energy in the IGM by photoionization through three channels.
The primary high velocity electron torn from hydrogen and helium atoms distributes its energy
by 1) collisional ionization, producing secondary electrons, 2) collisional excitation
of H and He and 3) Coulomb collision with thermal electrons. The fitting formula in
\citet{Shul85} is used to compute the fraction of secondary ionization and heating. 
These are taken into account when computing the evolution of the state of the IGM.

Then, it is legitimate to ask whether the lyman-$\alpha$ electrons resulting from the collisional
excitations are important for the Wouthuysen-Field effect. The simple answer is that, in our choice
of models, the energy emitted as X-ray is at most 10\% of the UV energy, itself $3$ times less than
the Lyman luminosity. Moreover at most 40\% of the X-ray energy is converted into excitations \citep{Shul85}, and
only $\sim 30$\% of the excitations result in a Lyman-alpha photon \citep{Prit06}. So, in the best case,
Lyman-$\alpha$ photons produced by X-ray represent only $0.3$\% of the photons produced directly by the sources.

\section{Results}

  \subsection{Ionization fraction}

\begin{figure}[h]
    \centering
    \resizebox{\hsize}{!}{\includegraphics{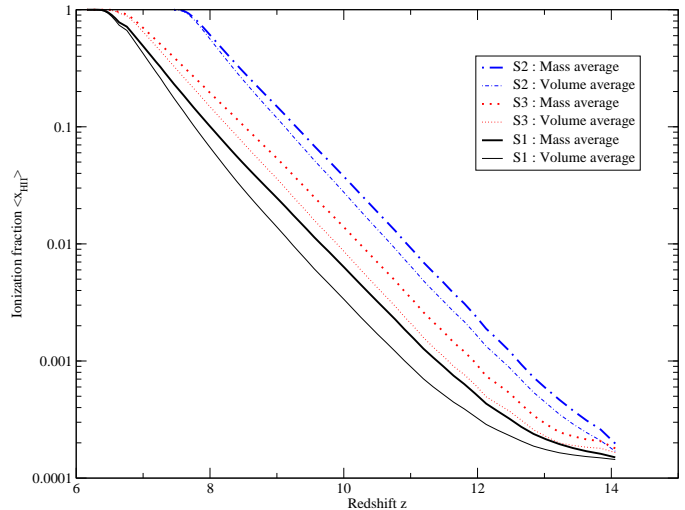}}
    \caption{Mass (thicker lines) and volume (thinner lines) weighted ionization fraction of hydrogen. S1 and 
S3 use Salpeter IMF and S2 use top-heavy IMF. S3 contains helium element.}
   \label{xh}
   \end{figure}

 The evolution of the averaged ionization fraction tells us about the global history of 
reionization. We plot the mass and volume weighted average ionization fraction in Fig.~\ref{xh}.
Including quasar (S4, S5, S6 and S7, not plotted) does not change the global evolution of the
ionization fraction much, because of its small fraction of the total luminosity. 
The total number of emitted photons is similar to S1. 
The S3 simulation has also the same number of emitted photons, but S3 reaches
the end of reionization a little bit earlier ($\Delta z \approx 0.25$) than S1.
Unlike other simulations, S3 contains helium which occupies
25\% of the IGM in mass. Including helium, the total number of atoms is reduced by
20\%. At the same time, unless X-ray are the dominant source of reionization, most of the helium 
is only ionized once  while $z>6$, due to the higher energy threshold for the secondary ionization($\rm {He}^{++}$).
 Therefore the number of emitted photons per baryon is higher in S3 than S1, and it
results in an earlier reionization.

On the other hand, S2 has the same number of photon absorbers as S1, but
the total number of emitted photons is much higher than in S1.
Using a top-heavy IMF, it produces 10 times more photons
(see Fig.~\ref{SED} and Tab.~\ref{luminosity_life}), and results in a $\Delta z \approx 1$
earlier reionization.

In all three cases, volume weighted values are less than mass weighted values,
since gas particles in dense regions around the sources are ionized first.
The volume occupied by each particle is estimated using the SPH smoothing length.

We computed the Thomson optical depth for all
simulations, the values are $\tau =$ 0.062, 0.076, 0.064 for S1, S2 and S3. The other simulations
(S4-S5) have the same $\tau$ as S1, since they follow the same evolution of ionization fraction.
They are somewhat lower than the Thompson optical depth derived from WMAP5 \citep{Hins09},
 $\tau =0.084\pm 0.016$,
only the S2 value is within 1$\sigma$ of the WMAP5 value. 
A variable escape fraction, decreasing with time, would allow the IGM to start
ionization earlier and increase $\tau$, without terminating ionization after z=6.

  \subsection{Gas temperature}
    The main goal of this study is to investigate the effect of inhomogeneous X-ray heating on the
21-cm signal. If the Ly-$\alpha$ coupling is sufficient, and X-rays can heat the gas
above the CMB temperature $T_\mathrm{CMB}$, the 21-cm signal will be observed in emission. However if 
the X-ray heating is not very effective, particularly during the early phase of the EoR,
we will observe the signal in absorption.

\begin{figure}[h]
    \centering
    \resizebox{\hsize}{!}{\includegraphics{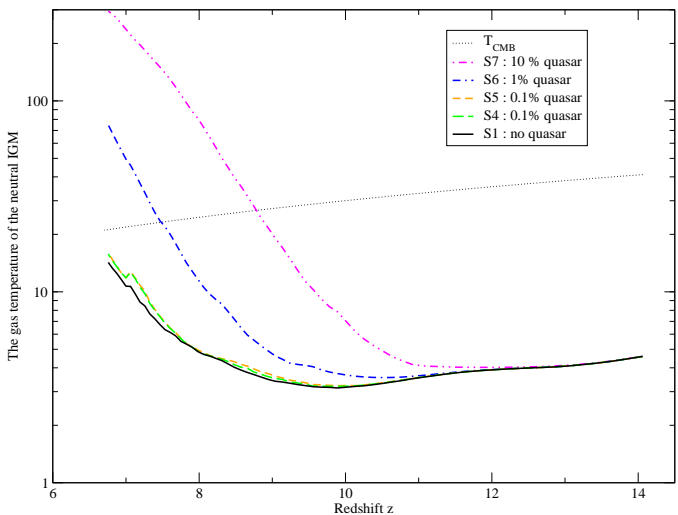}}
    \caption{The evolution of the gas temperature of the neutral IGM with redshift. 
The neutral gas is  chosen so that its ionization fraction is less than 0.01.}
   \label{Tk_evol}
   \end{figure}

We plot in Fig.~\ref{Tk_evol} the averaged gas temperature of the neutral IGM whose
ionization fraction $x_{HII}$ is less than 0.01.  
We chose the criterion of $x_{HII}<0.01$ for the following reasons. Once a gas particle is 10\% ionized,
it is heated by photoheating to a temperature of several thousand Kelvin. 
At redshift 10, the number of gas particles which have an 
ionization fraction between 0.01 and 0.1 is only 0.1\% of all the particles, but if we include these
particles, the average temperature increases from 2.94K to 5.41K. Therefore, we used the criterion $x_{HII}<0.01$ to evaluate
properly the average temperature of neutral regions, and verified that $x_{HII}<0.001$  gives
a very similar average temperature. We have checked that even for model S7 which has the highest level of X-rays, 
at $z > 7.5$, 99\%  of the
\textit{neutral} IGM has indeed an ionization fraction less than 1\%, so we have not excluded a significant fraction of the $21$ cm emitting
IGM from our average. 
This neutral gas is mostly located in the voids of the IGM.
 In fact, we have to consider Ly-$\alpha$ heating
as well as X-ray heating since a few K difference can reduce the intensity of $\delta T_b$
by up to 100 mK. 
 We recompute the gas temperature to include Ly-$\alpha$ heating 
 as a post-treatment using the formula
 from \citet{Furl06d}. This was detailed in Paper I. 	
The temperature of all simulations in Fig.~\ref{Tk_evol} decreases until $z\approx12$ 
because of the adiabatic expansion of the universe.
Then S7, which has the highest $L_{QSO}$, starts to increase first and reaches the CMB temperature at redshift $z\approx 8.8$.
Our fiducial model, S4, which contains 0.1\% of total energy as X-rays, shows
very little increase with respect to S1, a simulation without X-rays.
Even for S7, the gas temperature of neutral hydrogen in the void is still
under the $T_\mathrm{CMB}$ until $z\approx 8.8$. This means that
the X-ray heating needs time to heat the IGM above $T_\mathrm{CMB}$. Even with
a rather high level of X-ray, the absorption phase survives and produces 
greater brightness temperature fluctuations than the subsequent emission phase
(the delay in the absorption of X-ray connected to the long mean free path is 
partly responsible for this). It will be important to keep this result in mind
when choosing the design and observation strategies for the future instruments.

There is a large observational uncertainty in the mean and distribution
of the quasar spectral index $\alpha$. Our fiducial model assumes
$\alpha =1.6$ but we run a simulation (S5) with $\alpha =0.6$ for comparison.
The total emitted energy is fixed, but the S5 simulation has more energetic photons which
 penetrate the ionization front further than those of S4.
However, the difference of the gas temperature between two simulations is negligible.
The temperature of S5 is slightly higher than S4, but the difference is 
less than 0.5K at all redshifts. 

  We estimate that our 1\% model yields around $5$ times more X-rays than the fiducial model in \citet{Sant08}, although
the source formation modeling is quite different and the comparison is difficult (we do not use the dark matter halo mass
at all in computing the star formation rate).
However, comparing their plots of the average temperature and ionization fraction evolution with ours, we can
deduce that their average gas kinetic
temperature rises above the CMB temperature around ionization fraction $x_{\mathrm{HII}}=10\%$
while in our case the same event occurs at $x_{\mathrm{HII}}=15\%$. We
find several reasons for this apparent discrepancy. First we defined neutral IGM as $x_{\mathrm{HII}} < 0.01$ and use this to compute
the average gas temperature. Although this is not absolutely explicit in the paper, we believe they use $x_{\mathrm{HII}} < 0.5$ ,
thereby including warmer gas in the average. 
Then, they have a more extended reionization history, which reduces the effect
of the delay in the X-ray heating (see next section). Finally the initial X-ray heating is shifted to higher redshifts, when the
difference between the average neutral gas temperature and the CMB temperature is less.

  \subsection{Brightness temperature maps}

We have run Lyman-$\alpha$ simulations as a further post-treatment to obtain
the differential brightness temperature $\delta T_b$.
The $\delta T_b$ is determined by various elements, and it is expressed as  \citep{Mada97}:
\be
 \delta T_b \approx 28.1 \,\,{\mathrm{mK}}\,\,x_{\mathrm{HI}}\,\, (1+\delta) \left({1+z \over 10}\right)^{1 \over 2}\,\, {T_S -T_{\mathrm{CMB}} \over T_S},
\label{dTb_equation}
\ee
\noindent
where $\delta$ is the baryon over density, $T_s$ is the spin temperature, $T_\mathrm{CMB}$ is the CMB temperature, and $x_{\mathrm{HI}}$ is the neutral fraction.
The contribution of  the gradient of the proper velocity is not considered in this work.

The spin temperature $T_s$ can be computed with:
\be
T^{-1}_{S}= {T_{\mathrm{CMB}}+x_{\alpha}T^{-1}_c+x_cT^{-1}_K \over 1 + x_{\alpha} + x_c}
\label{T_spin}
\ee

\noindent
and

\be
x_{\alpha}={ 4 P_{\alpha} T_{\star} \over 27 A_{10} T_{\mathrm{CMB}} }  \qquad \mathrm{and} \qquad x_{c}={ C_{10} T_{\star} \over A_{10} T_{\mathrm{CMB}} }
\label{Ts_eq}
\ee

\noindent
 where $P_\alpha$ is the number of Lyman-$\alpha$ scatterings per atom per second, $A_{10}$ is the spontaneous emission coefficient of the $21$ cm hyperfine transition,
$T_{\star}$ is the excitation temperature of the $21$cm transition, and $C_{10}$ is the deexcitation rate via collisions. Details on deriving these relations and
computing $C_{10}$ can be found, e. g., in \citet{Furl06a}. The peculiar velocity gradients \citep{Bark05,Bhar04} 
is not considered in this work.

\begin{figure*}[th]
\centering
 \resizebox{0.85\hsize}{!}{\includegraphics{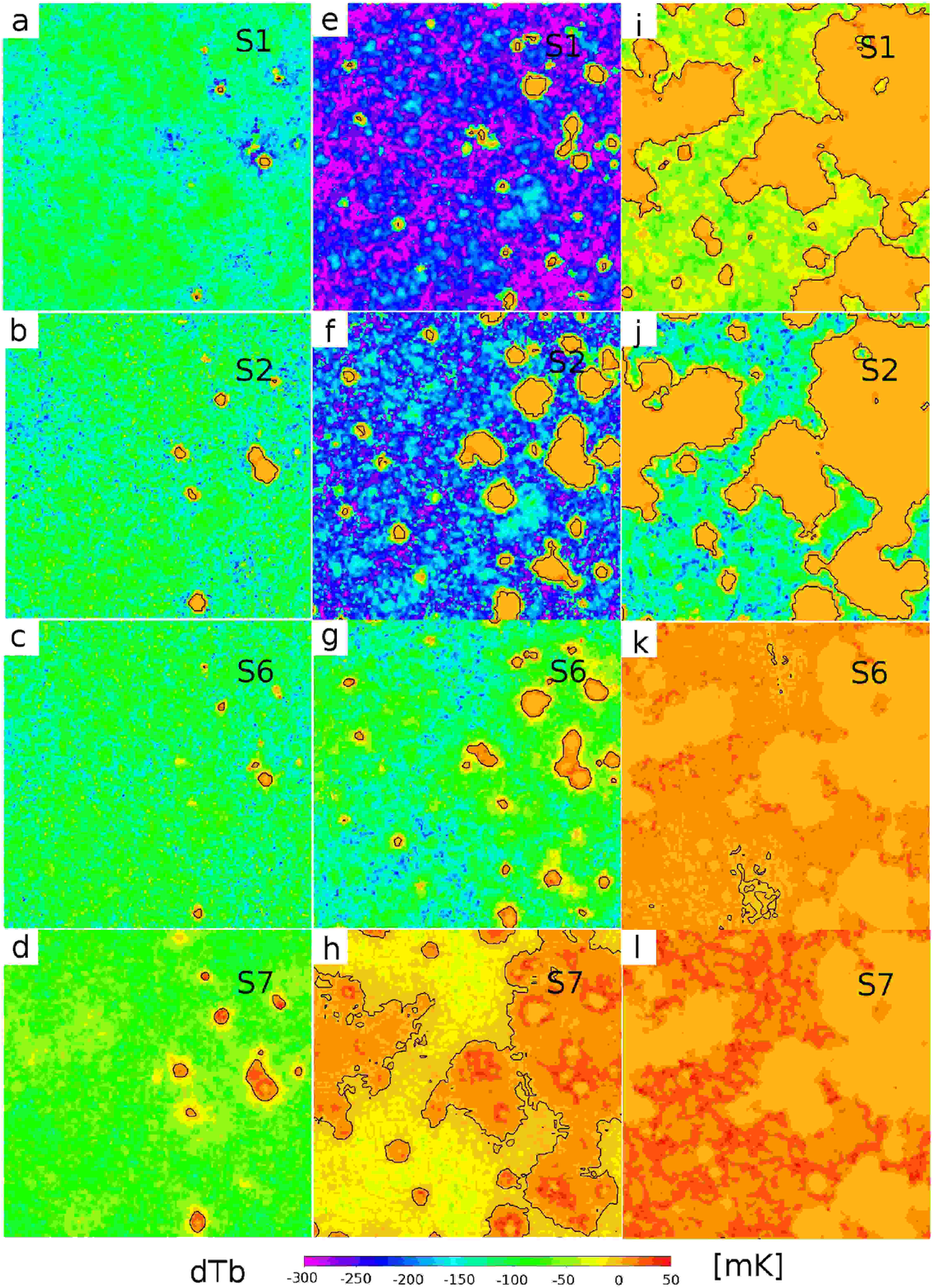}}
 \caption{Differential brightness temperature maps for different simulations. The thickness
of the slice is $\approx$ 2 Mpc. The maps in the left column
are when $\langle x_{\alpha} \rangle =1$, the middles when $\langle x_{\alpha} \rangle =10$
and the right when $\langle x_{\mathrm{HII}} \rangle =0.5$. The black contour separates
absorption and emission region. (l) shows no absorption region.}
\label{dTb}
 \end{figure*}

As we can see in eq. \ref{Ts_eq}, $T_S$ is coupled to the CMB temperature $T_\mathrm{CMB}$ by Thomson scattering of CMB
photons, and to the kinetic temperature of the gas $T_K$ by collisions and Ly-$\alpha$ pumping.
 Coupling by collisions is efficient only at $z > 20$, or in dense clumps, so Ly-$\alpha$ is the key coupling process in the diffuse IGM.
The $\delta T_b$ maps are a good way to see how these different
elements affect the signal. 
In Fig.~\ref{dTb} we show several $\delta T_b$ maps of the same slice from different radiative transfer simulations, S1, S2, S6 and S7.
S4 and S5, which sets the X-ray luminosity at 0.1 \% of the UV luminosity show
a trend very similar to S1 and are not plotted. 
The bandwidth of the slice is 0.1 MHz for all maps, which corresponds to 1.9 Mpc for the maps on the left column
(a)-(d), 1.8Mpc for (e)-(h) and 1.6 Mpc for (i)-(l).

 The left 
 four maps of $\delta T_b$ in  Fig.~\ref{dTb}, (a)-(d), are plotted when the mass averaged Ly-$\alpha$
coupling coefficient $x_{\alpha}$ is $\langle x_{\alpha} \rangle=1$. This value is interesting because in this moderate coupling regime,
fluctuations in the Ly-$\alpha$ local flux induce fluctuations in the brightness temperature, which is not the case  anymore when the coupling saturates.
The corresponding redshifts are $z=$ 10.50 for S2 and 10.13 for
the others. The corresponding averaged ionization fractions are 0.005, 0.018, 0.005 and 0.005 for (a)-(d).
Indeed, the averaged ionization fraction of S2 is higher than the others since it uses a harder spectrum.
The ratio of the integrated energy emitted in the Lyman band (Ly-$\alpha < E < $ Ly-limit)
with respect to the ionizing band, $\beta=E_{Lyman}/E_{ion}$, is three times less for S2 than for
the others. For a given number of emitted Ly-$\alpha$ photons, a harder spectrum produces a
larger number of UV ionizing photons, therefore S2 has a higher ionization fraction when 
$\langle x_{\alpha} \rangle=1$.
S1 shows a deeper absorption region around the ionized bubbles than S2, and it is also
due to the different ratio of the number photons in the Lyman band and the ionizing band.
In the case of S1, the ionized bubble is smaller than the highly Ly-$\alpha$ coupled region.
Since the kinetic temperature outside of ionized bubble is a few kelvin, which is lower than the
CMB temperature ($T_\mathrm{CMB}\approx 30 \,K$), the neutral hydrogen has a strong 21-cm absorption signal.
On the other hand, the highly Ly-$\alpha$ coupled regions in S2 mostly resides in the ionized bubbles, which
are bigger than in S1. S7 has almost the same averaged ionization fraction and ionized bubble size as S1, but
the gas around the ionized bubbles as well as in the void is heated by strong X-rays.
The signal is still in absorption
because the X-ray heating has not been able to raise the IGM temperature above 
$T_\mathrm{CMB}$, but the intensity is reduced.
Contrary to S1 and S2,  the neutral gas around the ionized bubbles produces a weaker signal 
than in the void, because the gas around the bubbles is more efficiently heated by X-rays. 
S6 shows an intermediate behavior between S1 and S7.

The four maps in the middle of Fig.~\ref{dTb}, (e)-(h), are for $\langle x_{\alpha} \rangle=10$.
The corresponding redshifts are $z=$ 9.03 for S2 and 8.57 for the others.
The averaged ionization fractions are 0.043, 0.141, 0.043 and 0.040 for (e)-(h).
These redshifts when $\langle x_{\alpha} \rangle=10$ are interesting
because the amplitude of  fluctuation $\delta T_b$ reaches a maximum.
 If we do not consider the effect of Ly-$\alpha$ coupling and assume $T_k \gg T_\mathrm{CMB}$, 
which  does not allow the signal in absorption, the largest fluctuations
would appear around $\langle x_{\mathrm{HII}} \rangle =0.5$ as noticed by \citet{Mell06} and \citet{Lidz07}, but
including the inhomogeneous
 Ly-$\alpha$ coupling and computing $T_K$ self-consistently, this maximum is shifted to
an earlier phase of reionization. 
The ionization fraction and bubble size in S2 are still
greater than in the other models, but the absorption intensity is lower than in S1.
Here is why. The evolution of kinetic temperature in the void regions is dominated by the
adiabatic cooling: the temperature drops as the expansion progresses.
 The kinetic temperature of S1 in the voids is lower than in 
S2 by 0.5 K due to the difference in redshift, which explains the stronger absorption intensity in S1.
 The neutral gas around ionized bubbles in 
S7 is heated by the high X-rays level above $T_\mathrm{CMB}$, and starts to produce the signal in emission.
The neutral gas in the voids is also affected by X-rays. However, it is not sufficiently heated yet
so that the signal turns everywhere from absorption to emission. Nevertheless the intensity of the signal
is reduced by the X-ray heating, and it shows the weakest signal among the four maps at $\langle x_{\alpha}\rangle$.

The four maps on the right of Fig.~\ref{dTb}, (i)-(l), are for $\langle x_{\mathrm{HII}} \rangle=0.5$.
The corresponding redshifts are $z=$ 7.68 for S2, 7.00 for S1 and S6, 6.93 for S7. The averaged
Ly-$\alpha$ coupling coefficients, $x_{\alpha}$, are 138.3, 34.9, 138.75 and 187.75 for (i)-(l).
Contrary to the above cases, the absorption intensity of S1 in the void region is 
weaker than that of S2. This is due to the Ly-$\alpha$ heating.
 Ly-$\alpha$ heating is negligible during the
 early phase, but the amount of Ly-$\alpha$ heating accumulated between
$z \sim 12$ and $z \sim 7$ can heat the gas
in the voids by several kelvins. 
In order to reach 50\% ionization, S1 produces a larger number of Ly-$\alpha$
photons, which propagate beyond the ionizing front. The $\langle x_{\alpha} \rangle$ 
of S1 is almost 4 times greater than that of S2, and the accumulated Ly-$\alpha$ heating
increases the kinetic temperature by 3-4 K more than in S2. 
The intensity in absorption is very sensitive to the value of kinetic temperature,
so this small amount of heating reduces the signal by up to 100 mK.
The X-ray heating in S7 is strong enough to heat all the gas above $T_\mathrm{CMB}$ at this
redshift, so we see the signal in emission everywhere. 
S6 shows intermediate features between S1 and S7, showing the weakest signal. Indeed,
the X-ray heating in S6 increased the gas temperature in the neutral voids just around
the $T_\mathrm{CMB}$, which is the transition phase from absorption to emission.

 \subsection{Power spectrum}

\begin{figure}[h]
    \centering
     \resizebox{0.9\hsize}{!}{\includegraphics{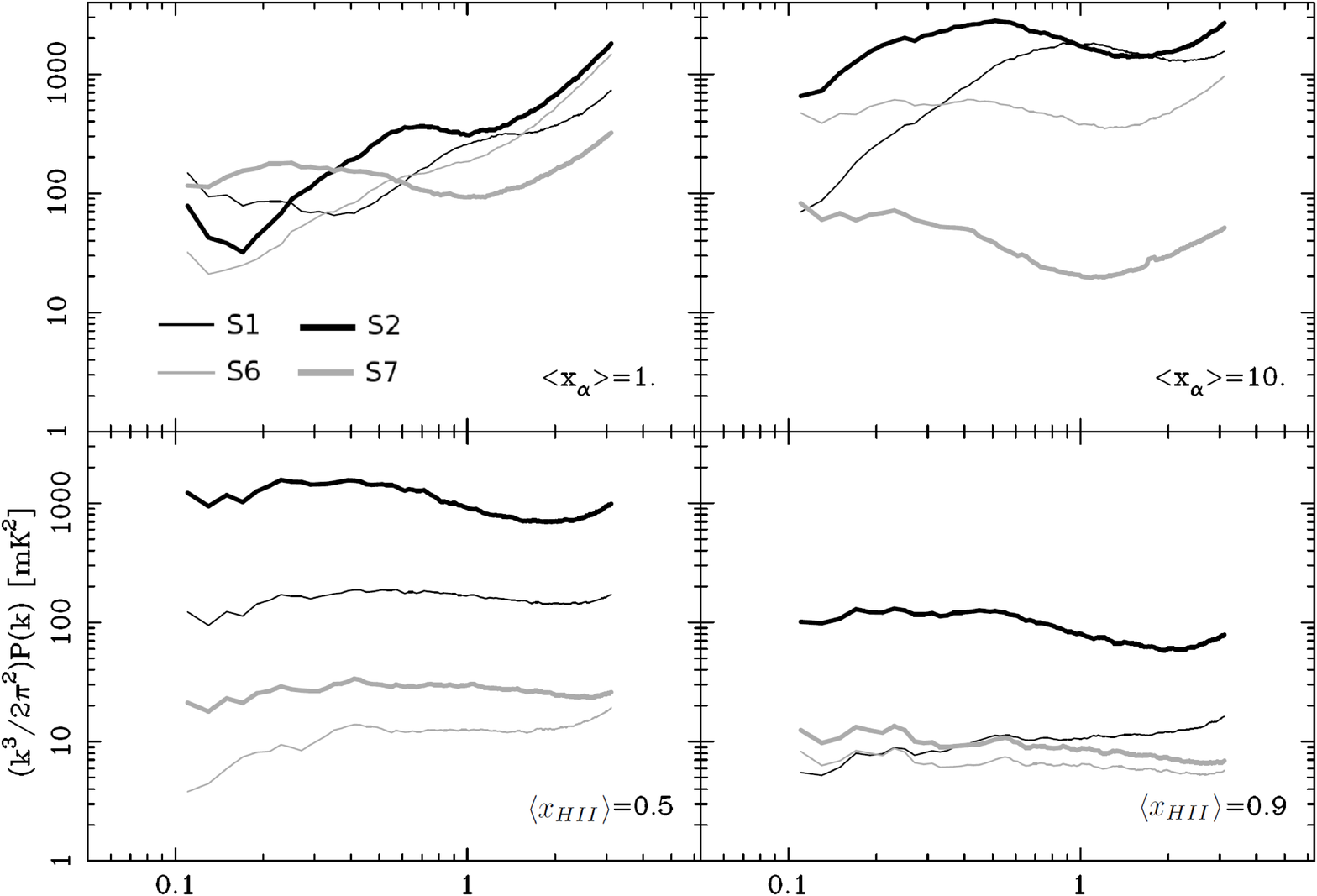}}
    \caption{Power spectrum evolution of the $\delta T_b$ from the S1 (thin black), 
      S2 (thick black), S6 (thin gray) and S7 (thick gray) simulations. S2 use top-heavy IMF
whereas the others use Salpeter IMF. S6 has 1\%  and S7 has
10\% of total luminosity for X-rays.}
   \label{power_panel}
   \end{figure}

 Fig.~\ref{power_panel} shows 3 dimensional power spectra of the brightness temperature
fluctuations for the S1, S2, S6 and S7 simulations. 
The power spectrum can be defined as the variance of the amplitude of the Fourier modes of the signal for a given
wavenumber modulus: 
\be
P(k)=\langle \hat{\delta T_b}(\mathrm{\bf k}) \hat{\delta T_b}^{\star}(-\mathrm{\bf k})\rangle.
\ee
We binned our modes with $\delta k=\frac{2\pi}{100}(\text{Mpc}/h)^{-1}$ and plotted the quantity 
$\Delta^2=k^3P(k)/2\pi^2$.
The power spectra of S4, S5 are not presented in 
Fig.~\ref{power_panel} since their patterns are similar to S1. 
During the early phase, when $\langle x_{\alpha} \rangle=1$, the amplitude of the powerspectrum of
 4 simulations
are similar. The spectra follow patterns similar to the power spectrum of S100 in Paper I (see Fig.15).
 Model S6, shows a spectrum similar to the fiducial model of \citet{Sant08}. The main difference
in shape appears at $k > 1 \mathrm{h}^{-1}.\mathrm{Mpc}$ in our model. This is possibly connected to the
fact that they assume a ${1 \over r^2}$ dependence of the Lyman-$\alpha$ flux, while at short distances from
the sources ( $< 10$ comoving Mpc), wings scattering effects produce a ${1 \over r^{7/3}}$ dependence \citep{Seme07}.
Also noticeable is the difference between S6 and S7. S7 is depleted at small scale, the strong X-ray heating damping
the absorption near the sources. On very large scale, however the already strong heating in S7 creates temperature
fluctuations which boost the S7 power spectrum.

When $\langle x_{\alpha} \rangle=10$, the power of both S6 and S7 decreases since X-ray heating
prevents a strong absorption signal. The strong X-ray heating of S7 increases the gas temperature
around $T_\mathrm{CMB}$, and it shows the smallest power.
However, the power of S6 falls down under the power of S7 when the hydrogen is 50 \% ionized.
At this redshift, the rising temperature of neutral gas reaches $T_\mathrm{CMB}$ in S6, while
it is already much greater than $T_\mathrm{CMB}$ in S7.  This is also visible in Fig.~\ref{dTb}.  
Later, when $\langle x_{\mathrm{HII}} \rangle=0.9$, all power spectra drop. Our S6 model agrees
quite well with \citet{Sant08} both in shape and amplitude for these two last stages. Indeed
both the effects of Lyman-$\alpha$ coupling and X-ray heating reach a saturation in the determination
of the brightness temperature, erasing the differences in our treatments.
S2 has the largest power over all scale when $\langle x_{\mathrm{HII}} \rangle=0.5$ and $\langle x_{\mathrm{HII}} \rangle=0.9$.
This is due to the near lack of Ly-$\alpha$ heating. Let us mention however that some sort of transition
to Pop II formation should have occurred by then, providing some level of Ly-$\alpha$ heating. So S2 is probably
not realistic during the late EoR.

 In brief, the 21-cm power spectra of our models vary in the 10 to 1000 mK$^2$ range, in broad agreement with \citet{Sant08}
who included the inhomogeneous X-ray and Ly-$\alpha$ effect on the signal in a semi-analytical way, with moderate discrepancies
at high-redshift and small scale due to wing effet in the Lyman-$\alpha$ radiative transfer. Quite logically our results differ
at high-redshift from \citep{Mell06b,Zahn07,Lidz07b,Mcqu06} who focused on the emission regime. These authors found a flattening of
the spectrum around $\langle x_{\mathrm{HII}} \rangle=0.5$. It is interesting to notice that in the case of a strong X-ray heating (model S7)
the spectrum is quite flat at all redshift (temperature fluctuation boost the power on large scales at high-redshift). 
In the future observations,
this would be a first clue of larger-than-expected contribution from X-ray sources.

We now plot the evolution of the power as a function
of redshift for 4 different $k$ values.
The evolution of the power spectrum with and without X-rays is very different.
S1 and S2, which do not have X-rays, show a \textbf{single} maximum on small scales ($k=1.00$ h/Mpc and $k=3.15$ h/Mpc)
around redshifts 8.5 and 9 which
correspond to the redshifts of $\langle x_{\alpha} \rangle =10$ for each simulation.
On large scales ($k=0.07$ h/Mpc and  $k=0.19$ h/Mpc) the power spectrum shows \textbf{two} local maxima on large scale.
The first peak is related to the Ly-$\alpha$ fluctuations. The $\delta T_b$ fluctuations
are dominated by Ly-$\alpha$ fluctuations at high-redshift, but it decreases when 
the Ly-$\alpha$ coupling saturates. Then it rises again. This time, 
the fluctuations are dominated by the fluctuations of ionization fraction. The second peak appears
at the redshift when $\langle x_{\mathrm{HII}} \rangle=0.5$ for each simulations. The
overall amplitude of S1 and S2 are similar, but the position of the local maximum peaks
of S2 are at higher redshift due to the faster reionization.  The key to the single-double peak
difference is that the contribution of the ionizing field fluctuation to the brightness temperature
power spectrum increases during reionization on large scale but not on small scales \citep{Ilie06b}.

\begin{figure}[h]
    \centering
     \resizebox{0.9\hsize}{!}{\includegraphics{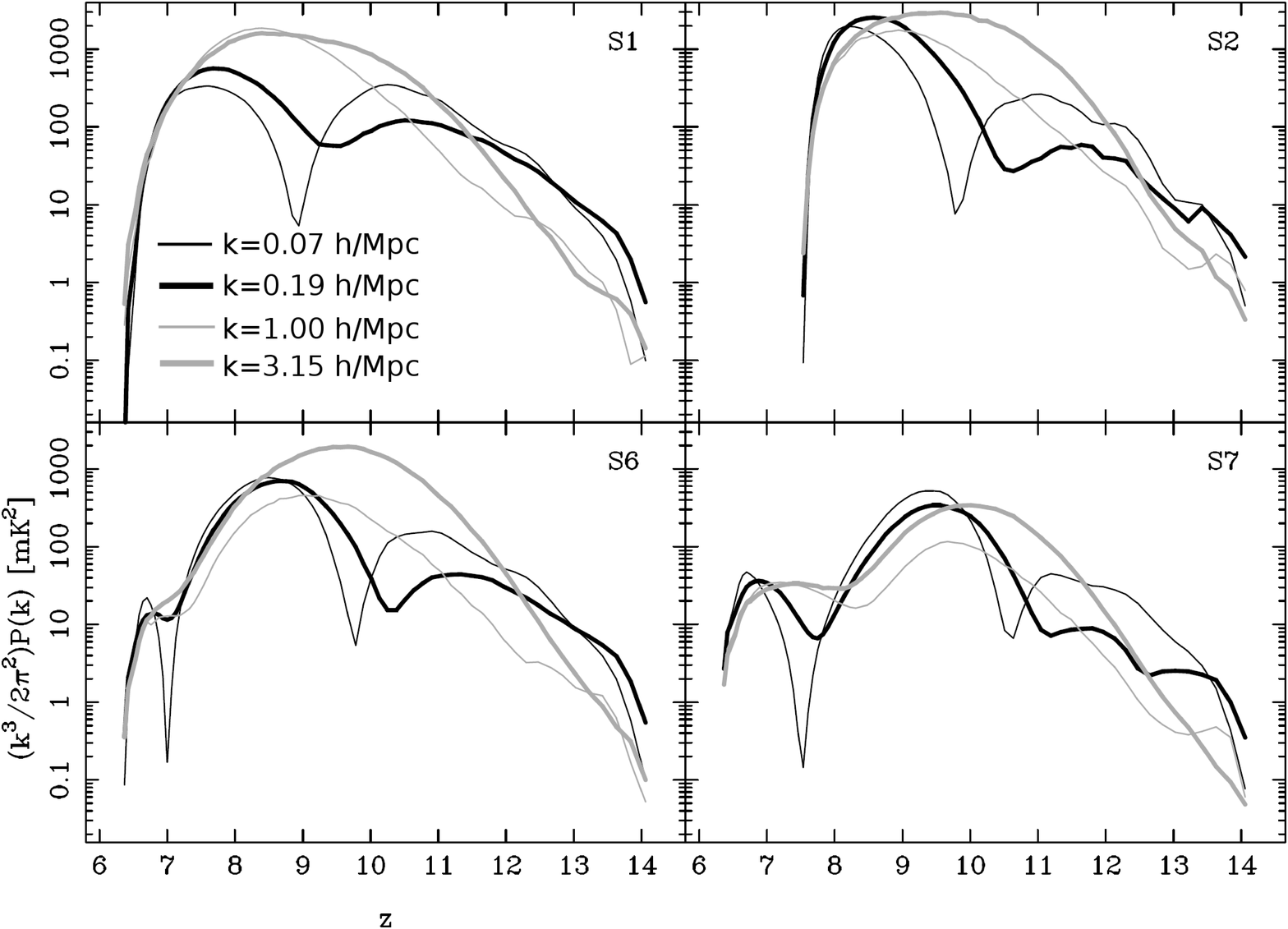}}
    \caption{Evolution of the brightness temperature power spectrum with redshift.
$k=0.07$ h/Mpc (thin black),  $k=0.19$ h/Mpc (thick black ), $k=1.00$ h/Mpc (thin gray) and $k=3.15$ h/Mpc (thick gray)}
   \label{power_evolution}
   \end{figure}

With X-rays (models S6 and S7), the evolution follows a different scenario. We find a pattern similar to
\citet{Sant08}.
On small scales (thin and thick gray in Fig.~\ref{power_evolution}), 
the intensity of the signal increases up to the maximum as the 
spin temperature couples to the kinetic temperature.
Then it decreases
during the absorption-emission transition. As the fluctuation due to the
ionization fraction comes to dominate, the power reincreases slightly or remains in 
a plateau until it drops at the end of reionization.
The evolution of the power on small scale does not show a marked minimum.  
The evolution of large scales (thin and thick gray in Fig.~\ref{power_evolution})
is the most interesting: it shows \textbf{three} maxima. From high-redshift to low redshift,
each peak corresponds to the period where the fluctuation of the Ly-$\alpha$ coupling,
 gas temperature and  ionization fraction dominate. 
There exists a deep suppression between the second and the third
peaks which does not appear without X-ray. It occurs when the X-ray heating raises the gas temperature
of the neutral IGM around $T_\mathrm{CMB}$, which dampens the signal.
The second minimum in S7 occurs earlier than in S6, since the stronger X-ray heating of S7
increase the gas temperature around $T_\mathrm{CMB}$ at a higher redshift.
We find a much narrower third peak in S4 (not plotted), which uses a 10 times weaker X-ray heating than S6.
The position and amplitude of the peaks as well as the width depend on the
intensity of X-ray heating. The width of the third bump ($6.5<z<7.5$) is the largest in S7 (with 10\% X-ray)
and the smallest or negligible in S4 (with 0.1\% X-ray).
The existence/position of this third peak and of
the second dip in the evolution of large scale power spectrum will be measurable by
LOFAR and SKA observations, and it will help us constrain the nature of the sources during the EoR.

  \subsection{Non-Gaussianity of the 21-cm signal}
The non-gaussianity of the 21-cm signal has been studied in previous works.
\citet{Ciar03c,Mell06b} shows the non-gaussianity of the 21-cm signal in numerical simulations by computing 
the Pixel Distribution Function (PDF). \citet{Ichi09} draws the history of reionization from the
measurement of the 21-cm PDF. \citet{Hark09} compute the skewness of the PDF and show 
how it could help in separating the
cosmological signals from the foregrounds.  However, all these works
model the signal in emission only.

We present the 21-cm PDF from our simulations in Fig.~\ref{pdf}, for several
representative redshifts. To obtain the 21-cm PDF, we sample $\delta T_b$ within 
a 1 $h^{-1}$Mpc resolution, an acceptable value for SKA.
The 21-cm PDF from our simulations is highly
non-Gaussian as expected, but  is also quite different from \citet{Ciar03c,Mell06b,Hark09} and \citet{Ichi09},
Our distributions extend to negative differential brightness temperature with a variety of shapes depending on the redshift.

The panel (a) of Fig.~\ref{pdf} is the 21-cm PDF at the beginning of reionization, $z=14.05$.
It is at the beginning of reionization that \citet{Ichi09} find a 21-cm PDF closest to
a Gaussian, but this is also when their model is the less relevant. In our case,
all signals are found in absorption and their distribution is peaked around
0mK, completely non-Gaussian.

We show the PDFs at $z=10.64$, still during the early EoR, in panel (b) of Fig.~\ref{pdf}.
The position of peaks are shifted around -100mK $\sim$ -50mK which means that 
the spin temperature of the particles is decoupled from $T_\mathrm{CMB}$ by Ly-$\alpha$ photons.
The PDF is much close to a gaussian distribution, extending to positive values (the smooth curves are the best gaussian fit to the PDFs). 
However, all of them are left skewed (toward negative temperature). The reason for this negative skewness is the same 
as the reason for the positive skewness in \citet{Mell06b,Ichi09,Hark09}: it is due to the signal from high density regions seen
in absorption for us and in emission for them.

Panel (c) in Fig.~\ref{pdf} shows the PDFs when $z=8.48$. Here we find a bimodal distribution with a plateau in the absorption
region between 100 mK and 0 mK, for models S1, S2 and S4. 
The left peaks of the PDFs moves also toward higher $\delta T_b$ with increasing heating efficiency.
In the case of S7, the left peak merges with the right one, and the form is very similar to a gaussian.

The panel (d) of Fig.~\ref{pdf} is plotted when the ionization fraction is 50\%. 
The width of the PDF of S1 and S4 is reduced because Ly-$\alpha$ heating is well advanced.
We find signals in emission in S6 and S7, since X-rays heat the gas in neutral regions 
above $T_\mathrm{CMB}$. Indeed, these PDF forms are similar to \citet{Ichi09}.
The PDFs of S6 and S7 could be fitted by the  Dirac-exponential-Gaussian distribution used by \citet{Ichi09}.
S2 has broad PDF still, since the Ly-$\alpha$ heating is 4 times lower than in the others and
it retains the signal in strong absorption.

It is interesting to note that the PDF always shows a spike around $\sim 0$ mK.
During the beginning of reionization, it is due to the large amount of neutral hydrogen whose 
spin temperature is still well coupled to the CMB temperature. As the reionization proceed, the spin temperature
is decoupled from the CMB and the number of pixels at $\sim 0$ mK decreases, but the peak grows again with the 
increasing contribution from completely ionized regions. This feature is interesting since interferometers such as
LOFAR or SKA only measure fluctuations in the signal and do not directly provide a zero point.

\citet{Ichi09} extract informations about the averaged
 ionization fraction from the 21-cm PDF. As could be expected
our PDFs converge with their result when the contribution from
X-ray sources is sufficient and the EoR somewhat advanced. What we find
is that a clear tracking of the nature of the ionizing source remains on
the PDF when the absorption phase is modeled. A strong X-ray contribution
produce unimodal PDFs while a weak X-ray contribution yield a bimodal PDF.

The evolution of the skewness is presented in Fig.~\ref{skewness}.
The skewness $\gamma$ is defined as
\begin{equation}
\gamma=\frac{\frac{1}{N}\Sigma _i (\delta T_{b}^i -\overline{ \delta T_{b}})^3 }
{[\frac{1}{N}\Sigma _i (\delta T_{b}^i -\overline{\delta T_{b}})^2]^{3/2}} \,\,,
\end{equation}
where N is total number of pixels in $\delta T_b$ data cube, $\delta T_{b}^i$ 
is $\delta T_b$ in $i^{th}$ pixel, and $\overline{\delta T_{b}}$ is the
average on the data cube.

At the beginning the skewness is highly negative for all simulations
as we can expect from the panel (a) and (b) of Fig.~\ref{pdf}. Then, in
all models, the skewness rises to a local positive maximum when
the average ionization fraction is a few percents.
It is interesting to notice that the skewness of all simulations
 close to zero again when the neutral fraction is about 0.3. While
this behavior could be used to provide a milestone of reionization, its
robustness should first be checked. We can notice that two of the three
models presented in \citet{Hark09} show the same behavior. Those are
however the two less detailed models.

Another interesting feature in Fig.~\ref{skewness} is that
the skewness of S7 has two local maxima while others do not.
Again, this could be used as a clue to a large contribution from
X-ray sources.

\begin{figure*}[tp]
\centering
\subfigure[]{\label{pdf35}\includegraphics[angle=-90,width=10cm]{Image/pdf35.ps}}
\subfigure[]{\label{pdf55}\includegraphics[angle=-90,width=10cm]{Image/pdf55.ps}}
\subfigure[]{\label{pdf75}\includegraphics[angle=-90,width=10cm]{Image/pdf75.ps}}
\subfigure[]{\label{pdf95}\includegraphics[angle=-90,width=10cm]{Image/pdf95.ps}}
\caption{The evolution of the 21-cm PDF. The redshifts are 14.05, 10.64 and 8.48 for
(a), (b), and (c). The PDFs of panel (d) is chosen so that the ionization fraction is 0.5.
The red curves on (b) are gaussian fits with the mean and variance of the PDFs. }
\label{pdf}
\end{figure*}

\begin{figure}[h]
    \centering
     \resizebox{\hsize}{!}{\includegraphics{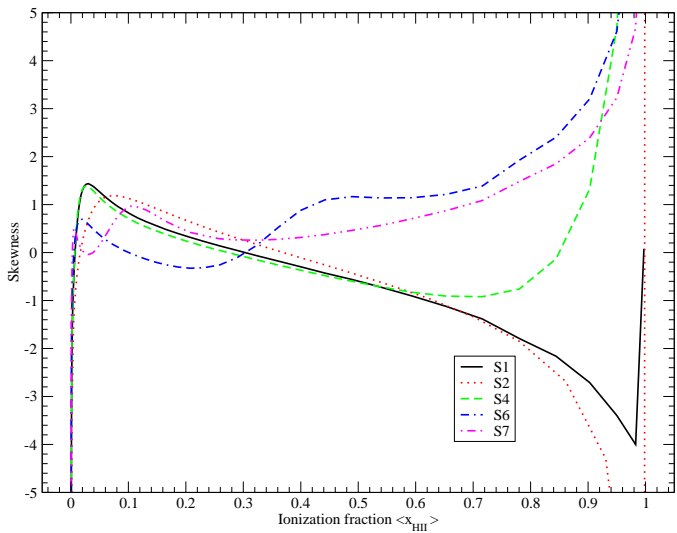}}
    \caption{Evolution of the skewness with the ionization fraction.}
   \label{skewness}
\end{figure}

\section{Conclusions}

We modeled the 21-cm signal during the EoR using numerical simulations
putting the emphasis on how various types of sources can affect the signal.
The numerical methods used in this work are similar to \citet{Baek09}.
The N-body and hydrodynamical simulations have been run with GADGET2 and post-processed with
UV continuum radiative transfer further processed with Ly-$\alpha$ transfer
using LICORICE, allowing us to model the signal in absorption. The main difference from the 
previous work is a more elaborated source model, including X-ray radiative transfer and  He chemistry. 

We have run 7 simulations to investigate the effect of different
IMFs, helium, different spectral indexes  and 
the different luminosities of X-rays sources.
The reference simulation in this work, S1, using only hydrogen and stellar type 
sources, reached the end of reionization at $z\approx 6.5$ and showed a strong absorption
signal until the end of reionization. 
Our top heavy IMF (model S2) produces $\sim 2.6$ times more ionizing photons than
the Salpeter IMF. S2 reached the end of reionization
earlier than the others by $\Delta z \approx 1$. In addition the different SED
changes the ratio of Ly-$\alpha$ and to ionizing UV photon numbers, and it slows down the
saturation of the Ly-$\alpha$ coupling and the heating by Lyman-$\alpha$ in the top heavy IMF case. 
This modifies the statistical properties of 21-cm signal.   

The simulation with helium, S3 also has a slightly earlier
reionization than the others since the number of emitted photons per baryon is higher.
Except for the slightly lower kinetic temperature in the bulk of ionized regions due to the
higher ionization potential than for hydrogen,
the properties of the 21-cm signal from S3 is similar to S1.
We chose QSO type sources with a power-law spectrum as X-ray sources in model S4 to S7.
The spectral index $\alpha$ has large observational uncertainty, so we used two different
spectral indexes. S4 and S5 have 0.1\% of the total luminosity in the  X-ray band.
 S5 uses $\alpha = 0.6$, while other simulations with X-rays use
$\alpha=1.6$. S5 showed very little difference on the gas temperature with respect to
 S4. 
S4, S6 and S7 have different  luminosities in the X-ray band, keeping
the same values for the other simulation parameters.
Using a stronger X-ray luminosity indeed increased the gas temperature
in the neutral hydrogen. Accordingly the 21-cm signal and its power spectra are
modified. 
We found an increase of a few kelvin for the neutral gas temperature in 
our fiducial model, S4, in which X-rays account for 0.1\% of the total emitted energy.
The 21-cm signal in S4 was similar to S1, showing the maximum intensity in
absorption,  $\sim 200$ mK, at $z \approx 9$. 
Stronger X-ray levels increase the gas temperature and reduce the intensity. We found that
in S6 and S7, which uses 1\% and 10\% of the total luminosity for X-rays, the absolute maximum intensity in
absorption decreases to $\sim 130$ mK and $\sim 80$ mK.
The 21-cm power spectrum of our work is greater by two or three orders of magnitude than
in works focusing on the emission regime \citep{Mell06b,Zahn07,Lidz07b,Mcqu06}. 
However, the results are in broad agreement with the work of
 \citet{Sant08}, who modeled absorption using semi-analytical methods for X-ray and Lyman-$\alpha$ transfer.
We noticed that the 21-cm fluctuation is dominated by Ly-$\alpha$ fluctuations during the early phase,
 X-rays later (or the gas temperature), and the ionization fraction at the end.
This is visible on  the evolution of the 21-cm power spectrum with redshift.
The 21-cm PDF of our work was different from other work, since
we do not assume that the spin temperature $T_s \gg T_\mathrm{CMB}$.

\textit{The first most important conclusion} from our work is that even including a higher than generally expected
level of X-ray, the absorption phase of the 21-cm survives. Its intensity and duration are reduced, but
the signal is still stronger than in the emission regime. Heating the IGM with X-rays takes time!

\textit{The second important result} is that we found three diagnostics which could be used in the analysis of future observations to
constrain the nature of the sources of reionization. $(i)$ The first and maybe the most robust is the
evolution with redshift of large scale modes ( $k \sim 0.1$ h/Mpc) of the powerspectrum. If reionization
is overwhelmingly powered by stars, this evolution should have one local minimum (two local maxima) . However, 
if the energy contribution of QSO is greater than $\sim 1\%$, a second local minimum (third maximum) appears.
The higher the X-ray level, the broader the third peak. $(ii)$ The second simple diagnostic is the bimodal aspect of the
PDF which disappears when the X-ray level rises above $1\%$ of the total ionizing luminosity. $(iii)$ Last is the redshift evolution
of the skewness of the 21-cm signal PDF. While all other models show a single local maximum at a few percent reionization, 
a very high level of X-rays ($> 10\%$ of the total ionizing luminosity) produces a second local maximum appear around $50\%$ reionization.

Modeling the sources in the simulation is complex. It involves taking the formation history, IMF, SED, life time, and more into account. Although
detailed models are desirable for the credibility of the results, we believe that the effect in the 21-cm signal can be bundled in 3 quantities.
The first is the efficiency: how many photons are produced by atoms locked into a star. This parameter must be calibrated to fit observational
constraint: end of reionization between redshift $6$ and $7$, and Thomson scattering optical depth in agreement with CMB experiment. The two other
quantities which contain most of the information are two box-averaged ratios: the energy emitted in the Lyman band  to the energy emitted in the ionizing band ratio 
and the same ionizing UV to X-ray  ratio. In this work we explored values of 0.32 (model S2) and 0.75 (all other models) for the former and $0.001$, $0.01$ and
$0.1$ for the latter. Once additional physics is included in the simulation and using a higher resolution to account for all the sources, it will be
interesting to explore the value of these quantities systematically.

We mentioned in the introduction that the minimum boxsize for reliable predictions of the signal is $100\,h^{-1}$Mpc.
It is important to realize that this value
(confirmed by emission regime simulations, e. g., \citealt{Ilie06b}) is estimated based on the clustering properties
of the sources and applies to the topology of the ionization field.
It may be underestimated when we study the early absorption regime, when only the highest
density peaks contain sources. Their distribution is the most
sensitive to possible non-gaussianities in the matter power spectrum. Moreover, they are
distant from each other and, consequently, produce large scale
fluctuations in the local flux of Lyman-$\alpha$ and X-ray photons. We intend to extend
our investigation to larger box sizes in a future work.

A few final words on additional physics not included in our model.
Shock heating from the cosmological structure formation is ignored,
but it could have the potential to affect the 21-cm signal by increasing the gas temperature
above the CMB temperature. However, it is not sure whether shocks are strong enough in the filaments
of the neutral regions to affect the 21-cm signal. Mini-halos ($\sim 10^4-10^8\,\,M_{\odot}$)
form very early during the EoR and are dense and warm enough from
shock heating during virialization to emit the 21 cm signal, but \citet{Furl06c}
find that the contribution of mini-halos will not dominate, because of the limited
resolution of the instrumentation. However, shock heating is worth investigating
with coupled radiative hydrodynamic simulations with higher mass resolution. Also worth
investigating is the effect of including higher Lyman lines in the radiative transfer.

\begin{acknowledgements}
This work was realized  in the context of the SKADS and LIDAU projects.
SKADS, the Design Study of the SKA project, financed by the FP6 of the
European Commission, the ANR project LIDAU, financed by the French
Ministry of Research.
\end{acknowledgements}

 \bibliographystyle{apj}
 \bibliography{ref}

\begin{appendix}
\section{Implemented method for UV and X-ray}
In this section, we explain the numerical methods that we use for radiative transfer with helium and X-ray.
A description of the main methods used in  LICORICE appears in \citet{Baek09b}
\footnote{Available at \url{http://aramis.obspm.fr/~baek/these.pdf}}.

\subsection{Hydrogen and helium ionization}

The radiation field is discretized into monochromatic photon packets which are emitted with
random direction and frequency (with appropriate distribution) from the point sources. 
Photon packets propagate through radiative
transfer cells and deposit photons and energy depending on the absorption probability of each cell. 
We need to compute absorption probabilities for each absorbers $\rm{H}^0$, $\rm He^0$ and $\rm He^+$.\footnote{
$\rm H^0$ denotes the neutral hydrogen and $\rm He^0$ denotes the neutral helium.}

The probabilities of the photon being absorbed by each elements are given by
\begin{equation}
 \cal{P}_{\rm{H}^0}=\frac{\tau_{\rm{H}^0}}{\tau_{\rm{H}^0}\,+\,\tau_{\rm{He}^0}\,+\,\tau_{\rm{He}^+}}(\rm{1} -\rm{e}^{-\tau})\,,
\end{equation}
\begin{equation}
 \cal{P}_{\rm{He}^+}=\frac{\tau_{\rm{He}^0}}{\tau_{\rm{H}^0}\,+\,\tau_{\rm{He}^0}\,+\,\tau_{\rm{He}^+}}(\rm{1} -\rm{e}^{-\tau})\,,
\end{equation}
\begin{equation}
 \cal{P}_{\rm{He}^0}=\frac{\tau_{\rm{He}^+}}{\tau_{\rm{H}^0}\,+\,\tau_{\rm{He}^0}\,+\,\tau_{\rm{He}^+}}(\rm{1} -\rm{e}^{-\tau})\,,
\end{equation}
where $\tau_{\rm{H}^0}$, $\tau_{\rm{He}^0}$ and $\tau_{\rm{H}^+}$ are the optical depths for 
$\rm{H}^0$, $\rm He^0$ and $\rm He^+$ in a given cell, and $\tau \equiv \tau_{\rm{H}^0}+ \tau_{\rm{He}^0}+\tau_{\rm{H}^+}$.
Therefore, the total absorption probability, $\mathcal{P}_{total}$, for a photon packet arriving in a cell of optical
depth $\tau$ is given by
\begin{equation}
\mathcal{P}_{total}(\tau)=\mathcal{P}_{\rm{H}^0}+\mathcal{P} _{\rm{He}^0}+\mathcal{P} _{\rm{He}^+}=1-\rm{e}^{-\tau}.
\end{equation}
For each cell, the optical depth $\tau$ is
\begin{eqnarray}
\tau &=& \tau_{\rm{H}^0} +\tau_{\rm{He}^0}+\tau_{\rm{He}^+} \nonumber \\
&=&\left[ \sigma_{\rm{H}^0}(\nu)n_{\rm{H}^0} +\sigma_{\rm{He}^0}(\nu)n_{\rm{He}^0}+\sigma_{\rm{He}^+}(\nu)n_{\rm{He}^+}\right]l,
\end{eqnarray}
where $\sigma _A$ is the photoionization cross-section for absorber $A\in \left\lbrace\rm{H}^0,\rm{He}^0,\rm{He}^+\right\rbrace $, $n^l_{A}$,
is number density in the cell and $l$ is the path length in a cell. We use photoionization cross section 
fits in \citet{Vern96}.

If $N_{\gamma}$ is the number of photons in a photon packet arriving in a cell, 
then the number of photons absorbed in a cell, $N_A$, is :
\begin{equation}
N_A=N_{\gamma}\mathcal{P}_{total}=N_{\gamma}(1-e^{-\tau}).
\end{equation}
The contribution of photoionization and photoheating to the evolution of the ionization fraction and the temperature within a time step $\Delta t$ is:
\begin{equation}
 \Delta x_{\rm{H}^+}=\frac{n_{\rm{H}}^0}{n_{\rm{H}}}\Gamma_{\rm{H}^0}\Delta t
= \frac{N_{\gamma} \mathcal{P}_{\rm{H}^0}}{N_{\rm{H}}} \,,
\end{equation}
\begin{equation}
 \Delta x_{\rm{He}^+}=\frac{n_{\rm{He}}^0}{n_{\rm{H}}}\Gamma_{\rm{He}^0}\Delta t
= \frac{N_{\gamma} \mathcal{P}_{\rm{He}^0}}{N_{\rm{He}}} \,,
\end{equation}
\begin{equation}
 \Delta x_{\rm{He}^{++}}=\frac{n_{\rm{He}}^+}{n_{\rm{H}}}\Gamma_{\rm{He}^+}\Delta t
= \frac{N_{\gamma} \mathcal{P}_{\rm{He}^+}}{N_{\rm{He}}} \,,
\end{equation}
\begin{eqnarray}
 \Delta T &=& \frac{2}{3k_B n} \Big\{- \frac{3}{2} k_BT\Delta n  + 
 n_{\gamma}\mathcal{P}_{\rm{H}^0}(h\nu -h\nu _{th,\rm{H}^0})\nonumber\\ 
&&+\, n_{\gamma}\mathcal{P}_{\rm{He}^0}(h\nu -h\nu _{th,\rm{He}^0}) \nonumber\\ 
&& +\, n_{\gamma}\mathcal{P}_{\rm{He}^+}(h\nu -h\nu _{th,\rm{He}^+})  \Big\} 
\end{eqnarray}
where $N_{\rm{H}}$ and $N_{\rm{He}}$ is the total number of hydrogen and helium in a cell, $h\nu_{th,A}$ are the ionization
potential of the recombined atom, $n_{\gamma}$ is the $N_{\gamma}$ in a unit volume. $\Gamma_A$ are the continuous photoionization rates used to actually integrate the evolution of th ionization fractions. Indeed
recombination, collisional ionization and radiative cooling are 
treated as continuous process, with integration time step $\Delta t^*$ much less than $\Delta t$ using the
following coupled equations 
\begin{eqnarray}
n_{\rm{H}}\frac{\rm{d}x_{\rm{H}^+}}{\rm{d}t}&=&\gamma _{\rm{H}^0}(T)n_{\rm{H}^0}n_e -\alpha _{\rm{H}^+}(T)n_{\rm{H}^+}n_e+\Gamma_{\rm{H}^0}n_{\rm{H}^0}\,,\nonumber\\
\nonumber\\
n_{\rm{He}}\frac{\rm{d}x_{\rm{He}^+}}{\rm{d}t}&=&\gamma _{\rm{He}^0}(T)n_{\rm{He}^0}n_e
-\gamma _{\rm{He}^+}(T)n_{\rm{He}^+}n_e\nonumber\\
& &\,-\,\alpha _{\rm{He}^+}(T)n_{\rm{He}^+}n_e\,\\\nonumber
& &+\alpha _{\rm{He}^{++}}(T)n_{\rm{He}^{++}}n_e +\Gamma_{\rm{He}^0}n_{\rm{He}^0}\,,\\\nonumber
\nonumber\\
n_{\rm{He}}\frac{\rm{d}x_{\rm{He}^{++}}}{\rm{d}t}&=&\gamma _{\rm{He}^+}(T)n_{\rm{He}^+}n_e \nonumber
-\alpha _{\rm{He}^{++}}(T)n_{\rm{He}^{++}}n_e \,\\\nonumber
& &+\Gamma_{\rm{He}^+}n_{\rm{He}^+},
\end{eqnarray}
where $\alpha_{I}$ and $\gamma_{A}$ indicate the recombination and collisional ionization coefficients and
$I\in {\left\lbrace\rm{H}^+,\rm{He}^+,\rm{He}^{++}\right\rbrace }$.
We use the values in \citet{Hui97} for the recombination coefficient and recombination cooling.
For collisional ionization coefficients and other radiative cooling we use the values in \citet{Cen92}.

The temperature is computed from the energy conservation equation. $E$ is the internal energy of the gas,
$E=\frac{3}{2}nk_B T$.
\begin{equation}
 \frac{\rm{d}E}{\rm{d}t}=\frac{\rm{d}}{\rm{d}t}\left( \frac{3}{2}nk_{\rm{B}}T \right)
=\mathcal{H}-\Lambda.
\end{equation}
$\mathcal{H}$ and $\Lambda$ are the heating and cooling function which account for the energy gained and lost
in a unit volume per unit time.

Each photon packet keeps propagating until it exits the simulation box (if we use free boundary condition) or
until the remaining photon content is much less than the initial photon content, $N_{\gamma}<10^{-p}N_{\gamma}^{initial}$.
We typically adopt $p=4$ for the UV continuum.

\section{Validation tests}

\subsection{Radiative Transfer Comparison Test}

 A cosmological radiative transfer code comparison project was performed in \citet{Ilie06}
trying to understand which algorithms (including various flavors of
ray-tracing and moment schemes) are suitable for a given non-trivial problem as well as
to validate each code by comparing the results with other codes. Five tests are run for radiative transfer in a static density field. We reproduced several of tests in \citet{Ilie06} with LICORICE.
LICORICE shows good agreements with other codes: results and comparisons with other codes are shown 
in \citet{Baek09b}.

Three additional tests are presented in \citet{Ilie09} for the coupled gas dynamical and radiative transfer evolution.
LICORICE directly participated in this second project and shows good agreements with the other codes.
 
\subsection{Comparison with CLOUDY I : helium}

An analytic solution to the radiative transfer evolution of an homogeneous medium around a single source 
exists only in the isothermal case with only hydrogen.
In order to validate helium ionization and spectrum hardening, we reproduced the
Str\"omgren sphere test in \citet{Mase03}, and compared our results with the 1-D 
radiative transfer code CLOUDY\footnote{\url{http://www.nublado.org/}} (C08 version of the code).

A point source, emitting as a black body at $T=60000$ K with ionizing luminosity
$L=10^{38}\, \rm erg\, s^{-1}$, is located at the center of the simulation box in a 
homogeneous density field with $n=1 \rm cm^{-3}$ composed of hydrogen (90\% by number) and
helium (10\% by number). The gas is initially completely neutral at a temperature
of $T=10^2$ K in the entire simulation box cube of $L_{box}=128$ pc. The comparison is 
performed at a time $t_s=6\times 10^5$ yr $\approx 5 t_{rec}^B$ (where $t_{rec}^B$ is
the characteristic time scale for hydrogen recombination).

The script used for CLOUDY is the following:

\vspace{0.5cm}
\begin{small}

\texttt{blackbody, T=60,000K}

\texttt{luminosity  38}

\texttt{radius 0.01 60 linear parsecs}

\texttt{hden -0.0457}

\texttt{abundances all -15}

\texttt{sphere static}

\texttt{element abundance helium -0.9542}

\texttt{punch element hydrogen "hydrogen\_caseA.dat"}

\texttt{punch element helium "helium\_caseA.dat"}

\texttt{punch temperature "temperature\_caseA.dat"}

\texttt{punch  continuum units eV  "spectrum\_60pc.dat"}

\end{small}

\vspace{0.5cm}

\begin{figure}[h]
    \centering
     \resizebox{\hsize}{!}{\includegraphics{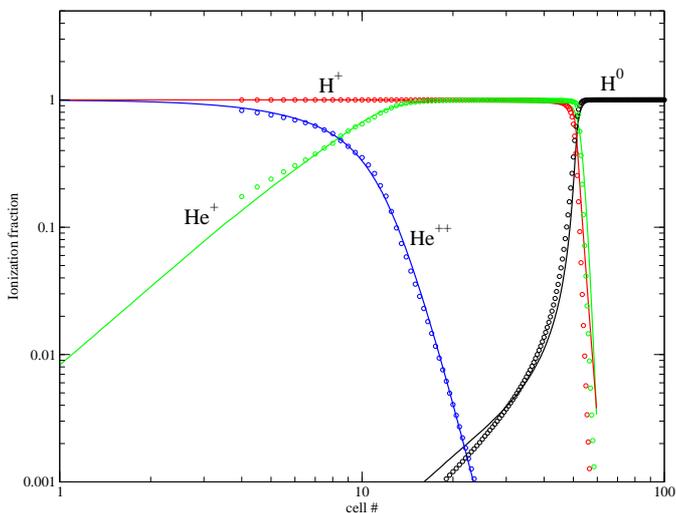}}
    \caption{Comparison of different ionization fractions between LICORICE (circles) and CLOUDY (solid lines),
as a function of distance from the point source in cell units (1 pc).}
\label{test1}
\end{figure}

\begin{figure}[h]
    \centering
     \resizebox{\hsize}{!}{\includegraphics{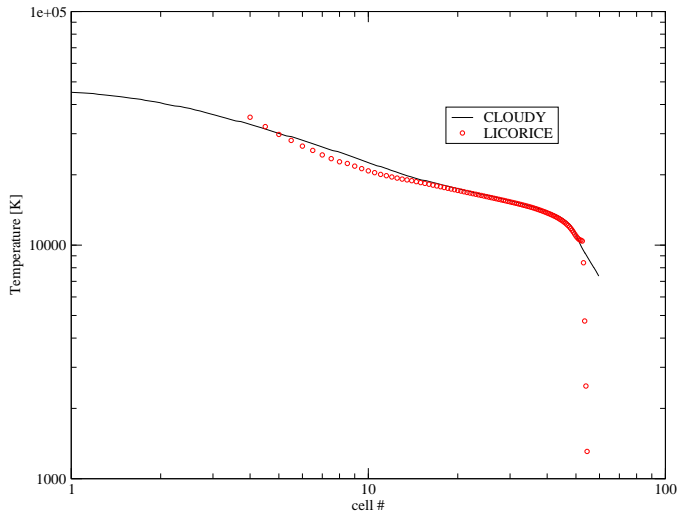}}
    \caption{Comparison of temperature distribution between LICORICE (circles) and CLOUDY (solid lines),
as a function of distance from the point source in cell units (1 pc). }
\label{test2}
\end{figure}

In Fig.\ref{test1} and Fig.\ref{test2}, the comparison between
CLOUDY (solid lines) and LICORICE (circles) is shown. The value of the different
physical quantities is plotted as a function of the distance from the source,
expressed in cell units, $\Delta x =1$pc. The points represent spherically
averaged LICORICE outputs. LICORICE shows a good agreement with CLOUDY.
The positions of various ionizing fronts agree within 1\%.
To obtain this good agreement we consider the effect of secondary ionization and heating
by fast electrons as CLOUDY does. The secondary ionization and heating fits in \citet{Shul85} are
only accurate for the primary photoelectrons higher than 100 eV, so we follow fits in \citet{Furl10} which
is valid for ones less than 100 eV.

The temperature profile shows also a good agreement, except for the warm tail
extending beyond the ionizing front for CLOUDY. This warm tail may originate from
heat conduction  which LICORICE does not implement.

\begin{figure}[h]
    \centering
     \resizebox{\hsize}{!}{\includegraphics{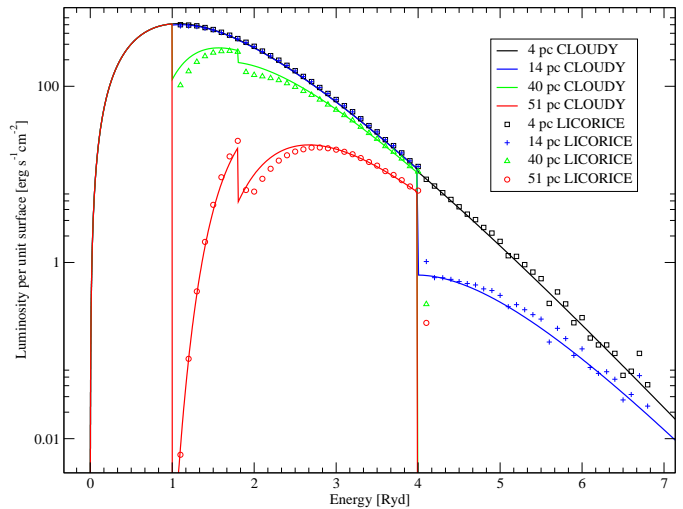}}
    \caption{The luminosity per unit surface at different distances. Comparison between CLOUDY (solid lines) and LICORICE (circles).}
\label{test3}
\end{figure}

Treating properly the spectrum hardening is an important issue in the radiative transfer with helium. 
Since each absorber ($\rm{H}^0$, $\rm He^0$ and $\rm He^+$) has a different ionizing potential, 
the emitted spectrum will be strongly depleted just above the different ionizing frequencies once it hits
a sufficient amount of absorbers.
We plot in Fig.\ref{test3} the luminosity per unit surface at different distances
and compare between CLOUDY (solid lines) and LICORICE (circles).
Our results agree well with the ones of CLOUDY. The noise at higher frequencies ($>$ 5 Ryd)
is due to the nature of the Monte Carlo method: the tails of the distributions are poorly sampled.
At 4 pc, all species are already ionized so the medium is transparent and the spectrum is close to the initial one.
At 14 pc, only photons with energy higher than 4 Ryd are absorbed to double ionize He.
Above 40 pc, we can observe also the depletion at $\sim 1.8$ Ryd, which corresponds to the single ionization of helium.

\subsection{Comparison with CLOUDY II : X-ray}

We have also performed a comparison test with CLOUDY to validate X-ray radiative transfer.
The geometrical initial set up is the same as in the test above.
Now a point source in the center emits X-ray photons with energy from
$100$ eV to $2$ keV, and a power-law spectrum with $\alpha=-1.6$ in Eq.\ref{power_index}.
The total luminosity is the same as in the helium test, $L=10^{38}\, \rm erg\, s^{-1}$.
The gas is composed of pure hydrogen, and completely neutral at the beginning.
The recombination time is proportional to the inverse of electron density. But instead of a sharp
ionization front we have a gradual decrease in the ionization fraction with increasing radius.
At the boundary of the box, the equilibrium ionization fraction is only $\sim 5 \%$ so the recombination
time is $\sim 20$  times large than in the inner region.
Consequently the integration with LICORICE was extended to $t=2500$ Myr.

The script used for CLOUDY is the following:
\vspace{0.5cm}
\begin{small}

\texttt{interpolate (1. -10.) (7.34 -10.) (7.35 0.) (147.35 -2.08) (147.4 -10.) (200. -10.)}

\texttt{luminosity  38}

\texttt{radius 0.1 100 linear parsecs}

\texttt{hden -0.0457}

\texttt{abundances all -15}

\texttt{sphere static}

\texttt{punch element hydrogen "hydrogen\_Xray.dat"}

\texttt{punch temperature "temperature\_Xray.dat"}

\end{small}

\vspace{0.5cm}

\begin{figure}[h]
    \centering
     \resizebox{\hsize}{!}{\includegraphics{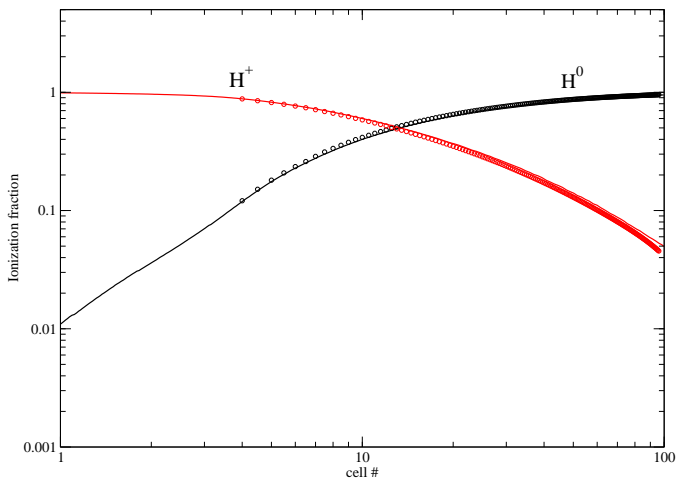}}
    \caption{Comparison of different ionization fractions between LICORICE (circles) and CLOUDY (solid lines),
as a function of distance from the point source in cell units (1 pc).}
\label{Xfraction}
\end{figure}

\begin{figure}[h]
    \centering
     \resizebox{\hsize}{!}{\includegraphics{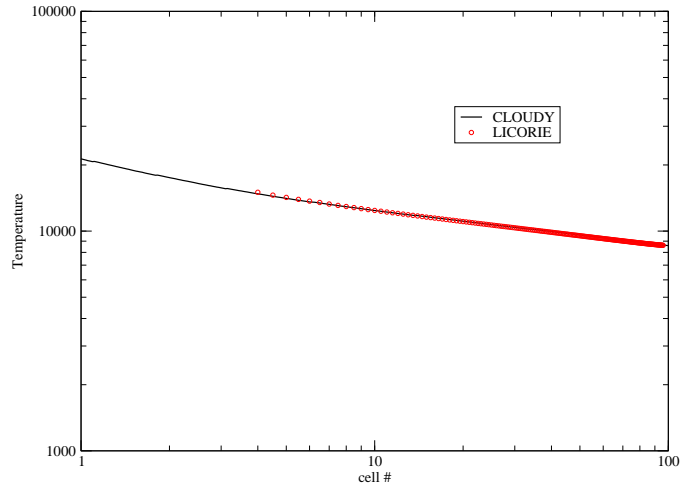}}
    \caption{Comparison of temperature distribution between LICORICE (circles) and CLOUDY (solid lines),
as a function of distance from the point source in cell units (1 pc). }
\label{Xtemperature}
\end{figure}

In Fig.\ref{Xfraction} and Fig.\ref{Xtemperature}, the comparison between
CLOUDY (solid lines) and LICORICE (circles) is shown. 
The value of the different
physical quantities is plotted as a function of the distance from the source,
expressed in cell units, $\Delta x =1$pc. The points represent spherically
averaged LICORICE outputs. LICORICE shows a very good agreement with CLOUDY. The (very) small difference
in the ionization fraction at large radii is due to the fact that we stopped the integration at 
$t=2500$ Myr, which is not much more than one local recombination time: the equilibrium is not fully
established. We could integrate for a longer time, but then the problem would only be shifted to larger radii and
smaller ionization fractions.

\section{Luminosity, SED, and typical life time of the sources}

Let us consider a source particle containing a population of stars. The number of stars $dN$ in a mass interval $dM$ is described by the IMF function $\xi(M)$:

$$dN=\xi(M)dM$$

The properties of a star of mass M are described by the bolometric luminosity $L(M)$, the effective blackbody temperature $T_{\mathrm{eff}}(M)$, and its life time $t_{\mathrm{life}}(M)$.
We can write the emissivity of a single star $\epsilon_\nu$ (erg.s$^{-1}$.Hz$^{-1}$):

$$ \epsilon_\nu(M) = L(M){ B_\nu(T_{\mathrm{eff}}(M)) \over \int_0^\infty B_\nu(T_{\mathrm{eff}}(M)) d\nu}\,, $$

\noindent
where $B_\nu$ is the Planck law. Then we can write the time dependent emissivity of the complete population $\epsilon_{\mathrm{tot}}$ as:

$$\epsilon_{\mathrm{tot}}(\nu,t)= {M_{\mathrm{source}}\over \int \xi(M) dM} \int H(t_{\mathrm{life}}(M)-t) \epsilon_\nu(M) \xi(M) dM $$

\noindent
were $H$ is the Heavy side step function and the integrals over the masses are bounded by the  mass interval in which we want to apply the IMF. With this quantity, we define a characteristic life-time in a spectral band $[\nu_1,\nu_2]$:

$$\tau_{\nu_1,\nu_2}= 2 {\int_{\nu_1}^{\nu_2} \int_0^\infty \epsilon_{\mathrm{tot}}(\nu,t)\, t \, dt d\nu \over \int_{\nu_1}^{\nu_2} \int_0^\infty \epsilon_{\mathrm{tot}}(\nu,t) \, dt \,d\nu}
$$
\noindent
If the total emissivity was a simple decaying exponential, this definition would give two characteristic decay times. We choose it as a typical time during which most of the energy has
been emitted. From this we can define the typical constant luminosity emitted in the band as:

$$ \overline{L}_{\nu_1,\nu_2}= {1 \over \tau_{\nu_1,\nu_2} } \int_{\nu_1}^{\nu_2} \int_0^\infty \epsilon_{\mathrm{tot}}(\nu,t) \, dt \, d\nu $$

\noindent
The value of $\tau_{\nu_1,\nu_2}$ and $ \overline{L}_{\nu_1,\nu_2}$ are computed in table 3 for different IMF in the Lyman and ionizing bands. Finally, we can compute the
time-averaged total emissivity:

$$ \overline{\epsilon}_{\mathrm{tot}}(\nu) \propto \int_0^\infty \epsilon_{\mathrm{tot}}(\nu,t) dt \propto \int t_{\mathrm{life}}(M) \epsilon_\nu(M) \xi(M
) dM $$

\noindent
This gives the constant SED to use during a characteristic life-time, and the normalization is given by the relation:

$$ \int_{\nu_1}^{\nu_2} \overline{\epsilon}_{\mathrm{tot}}(\nu) d\nu = \overline{L}_{\nu_1,\nu_2} $$

\end{appendix}

\end{document}